\documentclass[%
	usenat
]{aa}  

			\usepackage{amsmath}
			\usepackage{amssymb}
		
			\usepackage{graphicx}
		
			\usepackage{natbib}
			\bibpunct{(}{)}{;}{a}{}{,}	
	
			\usepackage{url}	

\title{}
\author{}

\begin{document}

   
		\title{AGN and QSOs in the eROSITA All-Sky Survey}
		
		\titlerunning{AGN and QSOs in the eRASS}
		
 		\subtitle{Part I: Statistical properties}
		
		\author{Alexander Kolodzig\inst{1}, Marat Gilfanov\inst{1,2}, Rashid Sunyaev\inst{1,2}, Sergey Sazonov\inst{2,3}, Marcella Brusa\inst{4,5,6}}
		
		\authorrunning{A. Kolodzig et al. (2013)}
		
		\institute{%
			Max-Planck-Institut f\"{u}r Astrophysik (MPA), Karl-Schwarzschild-Str. 1, D-85741 Garching, Germany, \email{alex@mpa-garching.mpg.de}
			\and
			Space Research Institute (IKI), Russian Academy of Sciences, Profsoyuznaya ul. 84/32, Moscow, 117997 Russia
			\and
			Moscow Institute of Physics and Technology, Institutsky per. 9, 141700 Dolgoprudny, Russia
			\and
			Max-Planck-Institut f\"{u}r extraterrestrische Physik (MPE), Giessenbachstrasse, Postfach 1312, D-85741 Garching, Germany
			\and
			Dipartimento di Fisica e Astronomia, Universit\'a di Bologna, viale Berti Pichat 6/2, 40127 Bologna Italy 
			\and
			INAF - Osservatorio Astronomico di Bologna, via Ranzani 1, 40127 Bologna, Italia
		}
		
		\date{Received 10.12.2012; accepted 15.08.2013} 
				
		\abstract
		%
		{
		The main element of the observing program of the Spectrum-Roentgen-Gamma orbital observatory is a four-year all-sky survey, in the course of which the entire sky will be scanned eight times.
		}
		%
		{ 
		We analyze the statistical properties of AGN (active galactic nuclei) and QSOs (quasars/quasi-stellar objects) that are expected to be detected in the course of the eROSITA all-sky survey (eRASS).
		}
		%
		{
		According to the currently planned survey strategy and based on the parameters of the Galactic and extragalactic X-ray background as well as on the results of the recent calculations of the eROSITA instrumental background, we computed a sensitivity map of the eRASS.
		Using the best available redshift-dependent AGN X-ray luminosity function (XLF), we computed various characteristics of the eRASS AGN sample, such as their luminosity- and redshift distributions, and the brightness distributions of their optical counterparts.
		}
		%
		{
		After four years of the survey, a sky-average sensitivity of about $1\times10^{-14}\;\mathrm{erg\;s^{-1}\,cm^{-2}}$ will be achieved in the $0.5-2.0\,\mathrm{keV}$ band. 
		With this sensitivity, eROSITA is expected to detect about $3$~million AGN on the extragalactic sky ($|b|>10\degr$).
		The median redshift of the eRASS AGN will be $z\approx1$ with approximately $40\,\%$ of the objects in the $z=1-2$ redshift range.
		About $10^4-10^5$ AGN are predicted beyond redshift $z=3$ and about $2\,000-30\,000$  AGN beyond redshift $z=4$,
		the exact numbers depend on the poorly known behavior of the AGN XLF in the high-redshift and luminosity regimes.
		Of the detected AGN, the brightest $10\,\%$ will be detected with more than $\approx38$ counts per point-spread-function (half-energy width), while the faintest $10\,\%$ will have fewer than $\approx9$ counts.
		The optical counterparts of approximately $95\,\%$ of the AGN will be brighter than $I_\mathrm{AB}=22.5\,\mathrm{mag}$.
		The planned scanning strategy will allow one to search for transient events on a timescale of half a year and a few hours with a $0.5-2.0\,\mathrm{keV}$ sensitivity of $\approx2\times10^{-14}$ to $\approx2\times10^{-13}\;\mathrm{erg\;s^{-1}\,cm^{-2}}$, respectively.
		}
		{}
		
		\keywords{%
			Surveys 
			-- X-rays: general
			-- X-rays: galaxies
			-- Quasars: general
			-- Galaxies: active
			-- Galaxies: luminosity function
			}
		
		\maketitle


		\section{Introduction} \label{sec:intro}
		
		
			\begin{table}[t]
				\renewcommand\arraystretch{1.3}		
				\renewcommand\tabcolsep{10pt}
				\caption{Predicted background count rates.}
				\label{tab:BKG}
				\centering 
				\begin{tabular}{l c c}	
					\hline
					\hline
					Energy band [keV]  & $0.5-2.0$ & $2-10$ \\
					\hline
					Particle        & 0.3 & 2.6 \\
					Galactic        & 1.8 & 0.0 \\
					Extragalactic & 1.9 & 0.5 \\
					\hline
					Total           & 4.0 & 3.1 \\
					\hline
				\end{tabular}
				\tablefoot{
				The count rates are given in units of $10^{-4}\;\mathrm{cts\;s^{-1}}$ per PSF HEW for seven telescopes.
				The extragalactic component accounts for unresolved sources only and at the average four-year survey sensitivity (Table~\ref{tab:Survey}).	
				}
			\end{table}
		
		
			\begin{table*}[t]
\renewcommand\arraystretch{1.3}		
\renewcommand\tabcolsep{10pt}
\caption{Characteristic average parameters of the eROSITA all-sky survey.}
\label{tab:Survey}
\centering 
\begin{tabular}{l | c c | c c | c c}
\hline
\hline
Survey duration		& \multicolumn{4}{c|}{4.0 years}		& \multicolumn{2}{c}{0.5 years}	\\
Sky region		& \multicolumn{2}{c|}{Extragalactic sky} & \multicolumn{2}{c|}{Ecliptic poles}	& \multicolumn{2}{c}{Extragalactic sky}	\\
Solid angle [$\mathrm{deg^2}$]	& \multicolumn{2}{c|}{$34\,100$ ($|b|>10\degr$)} & \multicolumn{2}{c|}{$90$}	& \multicolumn{2}{c}{$34\,100$ ($|b|>10\degr$)}	\\
Exposure time [sec] & \multicolumn{2}{c|}{$2\,000$} & \multicolumn{2}{c|}{$20\,000$} & \multicolumn{2}{c}{$260$} \\
\hline
Energy band [keV]       & $0.5-2.0$ & $2-10$ & $0.5-2.0$ & $2-10$  & $0.5-2.0$ & $2-10$ \\
\hline
Resolved extrag.\ CXB [$\%$]	& 31 & 6 & 53 & 17 & 12 & $\la 1$ \\
Background counts [cts/PSF]	& 0.8 & 0.6 & 6.7 & 6.1 & 0.1 & 0.1 \\
Source counts [cts/PSF]		& 7.6 & 6.8 & 16.5 & 15.6 & 4.4 & 3.9 \\
Sensitivity $\left<S_\mathrm{lim}\right>$ [$10^{-14}\,\mathrm{erg\,s^{-1}\,cm^{-2}}$] & $1.1$ & $18$ & $0.23$ & $4.2$ & $4.8$ & $80$ \\
Source density [$\mathrm{deg^{-2}}$] & 84 & 3.7 & 450 & 37 & 10.0 & 0.4 \\
Number of sources [$\times10^{3}$] & $2\,900$ & $130$ &  $41$ & $3.4$ & $340$ & $13$ \\
\hline
				\end{tabular}
			\end{table*}
		
		Large samples of X-ray detected active galactic nuclei (AGN) combined with follow-up optical data for identifying objects and their redshift determination are fundamental for understanding AGN evolution and the growth of supermassive black holes (SMBHs) with cosmic time.
		These samples are constructed in various extragalactic X-ray surveys spanning from wide-shallow to narrow-deep surveys.
		Many of these have been conducted in the past decade  with the Chandra and XMM-Newton X-ray observatories, which were instrumental in understanding the cosmic X-ray background and evolution of AGN at low- and high redhifts \citep{Brandt2005}.
While Chandra and XMM-Newton have now surveyed several hundreds of square degrees (e.g. $\sim360\;\mathrm{deg^2}$ of the 2XMM-catalog, \citealt{2XMM}),
the sensitivity of the archival observations is far from be homogeneous.
Moreover, the sky area covered by largest contiguous surveys did not exceed several (e.g. XMM-COSMOS: \citealt{XMMCOSMOS}; XBootes: \citealt{Murray2005}) to several tens of square degrees (e.g. XMM-LSS: \citealt{XMMLSS}; XWAS: \citealt{XWAS}; XXL: \citealt{XXL}),
which in particular limited our knowledge of the evolution and clustering properties of the most luminous AGN severely.
		The most recent \emph{all-sky survey} \citep{Voges1999} in the X-ray band was performed by ROSAT%
			\footnote{\url{http://www2011.mpe.mpg.de/xray/wave/rosat/}\label{fn:ROSAT}}
		\citep{Truemper1993} more than two decades ago, creating an increasing demand for an all-sky survey to be conducted by the new generation of X-ray telescopes.
	
		The eROSITA\footnote{\url{http://www.mpe.mpg.de/eROSITA}} telescope (extended ROentgen Survey with an Imaging Telescope Array) will be able to statisfy this demand.
		It is the main instrument aboard the Russian Spektrum-Roentgen-Gamma satellite\footnote{\url{http://hea.iki.rssi.ru/SRG}}, which is scheduled for launch in 2014. 
		Its main science goals are cosmological studies of clusters of galaxies and AGN, with the aim of constraining the nature of dark matter and dark energy.
		To achieve these goals, eROSITA will perform an all-sky survey (eRASS) during the first four years of its operation, followed by a phase of pointed observations.
		The main mission parameters and the telescope are described in the eROSITA Science Book \citep[hereafter SB]{eROSITA.SB}.
		
		In this study we explore the main statistical properties of the AGN sample that is expected to be detected in the course of the eRASS, including its luminosity- and redshift distributions.
		This will help to understand the capabilities of the eROSITA mission and, potentially, to tune the survey strategy and its parameters.
						
		We use the following cosmological parameters throughout: $H_0 = 70.0\;\mathrm{km\,s^{-1}\,Mpc^{-1}}$, $\Omega_m = 0.30$, $\Omega_\Lambda = 0.70$, $\Omega_k = 0$.
		These values are commonly used for XLF modeling of AGN.
		We use the decimal logarithm throughout.
		The calculations are performed for two energy bands -- soft ($0.5-2.0\;\mathrm{keV}$) and hard ($2.0-10.0\;\mathrm{keV}$).
		In computing count rates we used the most recent response matrix of eROSITA, \texttt{erosita\_iv\_7telfov\_ff.rsp}%
			\footnote{\url{http://www2011.mpe.mpg.de/erosita/response/}\label{fn:RSP}}.
		As is appropriate for the all-sky survey data analysis, this response matrix is averaged over the field-of-viewand scaled to 7 telescopes.
		In this work, we assume that the data from the entire survey of eROSITA is available for analysis.
		
		\section{Sensitivity} \label{sec:Sens}

		The point-source detection sensitivity of eROSITA in the all-sky survey was discussed in detail by \citet{Prokopenko2009}.
		Since then, the spacecraft orbit has been changed to the L2 orbit and detailed calculations of the instrumental background became available.
		We therefore update the calculations of \citet{Prokopenko2009} below.
		We first compute the sky average values (Sect.~\ref{ssec:BKG}-\ref{ssec:ConL})
		and then calculate a more realistic  sensitivity map (Sect.~\ref{ssec:SensMap}),
		which is used for our eRASS AGN calculations.

		\subsection{Instrumental and cosmic background}	\label{ssec:BKG}	
		
		
			\begin{figure*} 
				\centering
				\includegraphics[width=17.5cm]{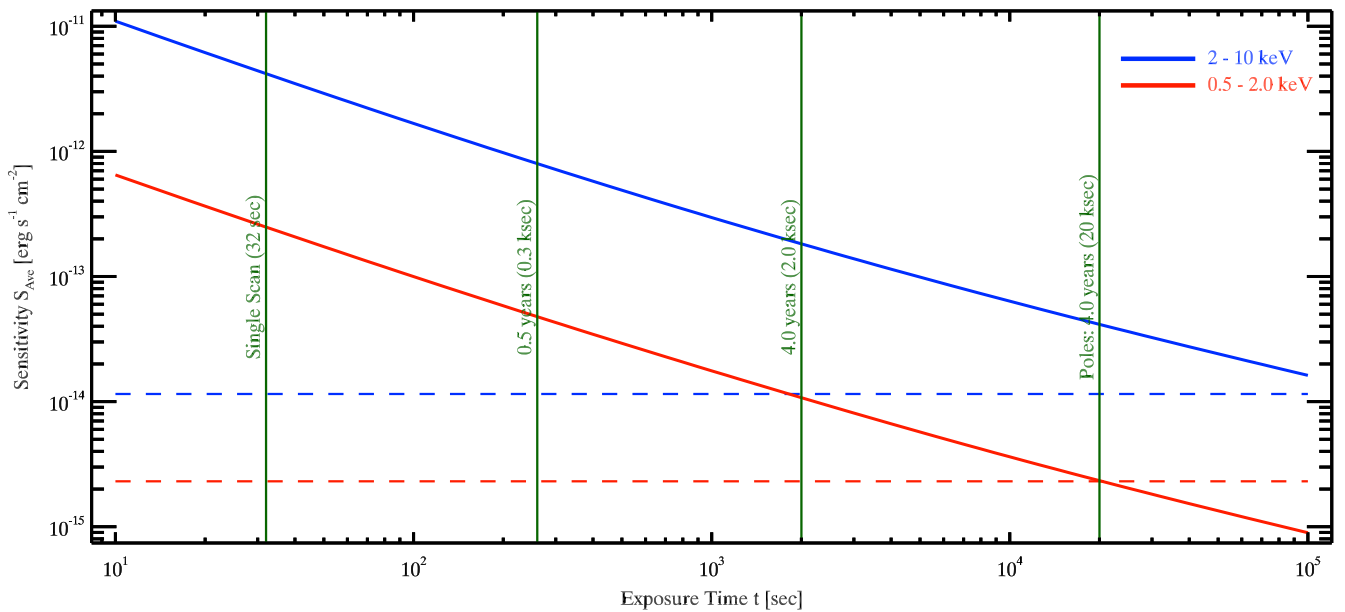}
				\caption{\label{fig:sens}
					Average point-source detection sensitivity of the eROSITA telescope as a function of the exposure time in the soft (\emph{red}) and hard (\emph{blue}) energy bands for the extragalactic sky.
					The horizontal dashed lines are the confusion limits for one source per $40$ telescope beams (PSF HEW).
					The vertical line on the left shows the exposure time for a single scan at $100\,\%$ observing efficiency, the other vertical lines indicate the average exposure times for different survey durations and at the ecliptic poles (for the soft-band confusion limit) at $80\,\%$ observing efficiency.
				}
			\end{figure*}
		
		The eROSITA background is dominated by the photon background below $\sim2$ keV and by the particle component above this energy.
		In the subsequent calculations we used the field-of-view-averaged angular resolution (PSF\footnote{PSF - Point-spread-function} HEW\footnote{HEW - Half-energy width, alias half-power diameter (HPD)})
		of $30^{\prime\prime}$ diameter for the soft band and $40^{\prime\prime}$ diameter in the hard band \citetext{Friedrich et al., priv.\ comm.}.
		The corresponding PSF HEW areas are $\approx707$ and $\approx1\,257\;\mathrm{arcsec^2}$, respectively.
		For the eRosita focal length, $1\;\mathrm{mm}$ on the detector corresponds to $\approx128.8\;\mathrm{arcsec}$.

		The instrumental non X-ray (particle) background spectrum is nearly flat in the counts space, with a normalization of $\approx6.1\times10^{-3}\;\mathrm{cts\;s^{-1}\,cm^{-2}\,keV^{-1}}$ \citep{Perinati2012}. 
		Within the PSF HEW it produces count rates
		of $\approx2.7\times10^{-5}$ and $\approx2.6\times10^{-4}\;\mathrm{cts\;s^{-1}}$ for the soft and hard band for seven telescopes.
		Solar-activity-induced background events are not taken into account in this calculation, they are roughly accounted for via the observing efficiency introduced in the survey exposure time calculations.
		These numbers are the result of purely theoretical calculations and there are no direct measurements of the real background of an X-ray detector in the L2 orbit.
		Therefore the above numbers may have to be revised after the eROSITA launch.
		
		The X-ray photon background (CXB) has two components \citep{Lumb2002}:
		(1) the truly diffuse Galactic background of local ionized ISM\footnote{ISM - Interstellar medium} with a soft thermal spectrum, and
		(2) the hard power-law extragalactic CXB component.

		To estimate the contribution of the ionized ISM emission we used the spectral fits from \citet[Table~3]{Lumb2002}
		and obtained a count rate of $\approx1.8\times10^{-4}\;\mathrm{cts\;s^{-1}}$ within the PSF HEW for the soft band, assuming the same Galactic absorption ($N_\mathrm{H}=1.7\times10^{20}\;\mathrm{cm^{-2}}$) and solar abundance \citep{Anders1989} as \citet{Lumb2002}.
		Its contribution to the hard band can be neglected.		
		As a caveat we note that the \citet{Lumb2002} analysis was based on the XMM-Newton observations of several blank fields located at high Galactic latitudes.
		Therefore these numbers should be considered as approximate, because they  do not  account for inhomogeneity the of the Galactic background radiation.

		For the extragalactic CXB component we assumed a power-law spectrum with a photon index of $\Gamma=1.42$ \citep[Table~3]{Lumb2002}.
		The power-law was normalized to the extragalactic CXB flux of $7.53\times10^{-12}$ and $2.02\times10^{-11}\;\mathrm{erg\;s^{-1}\,cm^{-2}\,deg^{-2}}$ for the soft and hard band \citep{Moretti2003}%
			\footnote{%
				Strictly speaking, these flux values correspond to a slightly steeper slope ($\approx1.45$) of the CXB spectrum than the conventional value of $1.42$.
				This discrepancy reflects the uncertainty in the absolute CXB flux determinations.
				We nevertheless used them for consistency with the resolved fraction calculations below.%
			}.
		We furthermore assumed Galactic absorption of $N_\mathrm{H}=6\times10^{20}\;\mathrm{cm^{-2}}$ corresponding to the arithmetic mean of the $N_\mathrm{H}$-map of \citet{Kalberla2005} for the extragalactic sky ($|b|>10\degr$).
		With these parameters, the average count rate of extragalactic CXB within the PSF HEW (seven telescopes) is $\approx2.8\times10^{-4}$
		and $\approx5.8\times10^{-5}\;\mathrm{cts\;s^{-1}}$ in the soft and hard band, respectively.
		
		In computing the contribution of extragalactic sources, one needs to take into account that a fraction of the background AGN will be resolved by eROSITA.
		Therefore, these sources will not contribute to the unresolved image background, which will affectg the point-source detection sensitivity.
		This effect reduces of the extragalactic background count rate.
		At the average four year survey sensitivity of eROSITA (see Sect.~\ref{ssec:AveSens}) the resolved extragalactic CXB fraction%
			\footnote{
			Note that Fig.~5 in \citet{Moretti2003} gives resolved fractions of $\sim50\,\%$ and $\sim10\,\%$, respectively.
			This difference also reflects the uncertainties of the CXB measurements.
			}
		achieves $\approx31\,\%$ in the soft and $\approx6\,\%$ in the hard band.
		The fractions were calculated using the number counts of \citet[hereafter G08]{Georgakakis2008} and the extragalactic CXB flux of \citet{Moretti2003}.
		Thus, the final values of the average count rate of the unresolved CXB emission within the PSF HEW is $\approx1.9\times10^{-4}$ and $\approx5.4\times10^{-5}\;\mathrm{cts\;s^{-1}}$ in the soft and hard band, respectively.

		To examine the background model, we used ROSAT all-sky maps of  diffuse X-ray emission \citep{Snowden1997}.
		We combined  PSPC  maps%
			\footnote{\url{http://www.xray.mpe.mpg.de/rosat/survey/sxrb/12/ass.html}}
		in the energy bands from \emph{R4} to \emph{R7}  \citep[see][Table~1]{Snowden1997} to cover the $0.44-2.04\;\mathrm{keV}$ range, which presents a reasonable approximation of the eROSITA soft band.
		The  median count rate on the combined map is  $\approx0.87\;\mathrm{PSPC~cts\;s^{-1}\,deg^{-2}}$  for the extragalactic sky ($|b|>10\degr$).
		A convolution of our X-ray background model with the ROSAT PSPC response matrix 
		(\texttt{pspcb\_gain2\_256.rsp}\footnote{\url{ftp://legacy.gsfc.nasa.gov/caldb/data/rosat/pspc/cpf/matrices/}})
		gives a $0.44-2.04\;\mathrm{keV}$ count rate of $\approx0.99\;\mathrm{cts\;s^{-1}\,deg^{-2}}$, which is sufficiently close to the median value in  the ROSAT map.
		ROSAT maps show moderate brightness variations in which $90\,\%$ of the extragalactic sky  count rate are between $\approx0.34$ and $\approx2.17\;\mathrm{PSPC~cts\;s^{-1}\,deg^{-2}}$.
		Variations of the background count rate in this range amount to $\sim30\%$ variations in the sensitivity.
		
		The contributions of different background components are summarized in Table~\ref{tab:BKG}.
		They are consistent within $\approx10\,\%$ with the numbers in the SB.
		The difference in the soft band comes from the slight difference in the normalization of the extragalactic component.
		The change in the hard band appears because we used the results of the updated particle background calculations of \citet{Perinati2012} instead of those of  \citet{Tenzer2010} which were used in the SB.

		\subsection{Average exposure and sensitivity} \label{ssec:AveSens}

		With the average background count rates we computed the point-source detection sensitivity of eROSITA as a function of the exposure time, which is shown in Fig.~\ref{fig:sens}.
		In this computation we assumed a Poissonian distribution of counts and demanded no more than $200$ false point-source detections for the entire sky.
		This corresponds to one false detection in $\approx250$ fields of view ($\approx210\;\mathrm{deg^2}$).
		For a Gaussian distribution, this false-detection rate is equivalent to an $\approx5.0\sigma$ 
		confidence level in one trial.
		In converting the count rates into flux we assumed an absorbed power-law spectrum with a photon index $\Gamma=1.9$ and a sky-averaged Galactic absorption of $N_\mathrm{H}=6\times10^{20}\;\mathrm{cm^{-2}}$.
		We also took into account that only half of the source counts are contained within the PSF HEW.

		The sky-averaged exposure time of the survey is
			\begin{align} \label{eq:ET}
				t_\mathrm{exp} \approx 2.0
					\; \left(\dfrac{t_\mathrm{survey}}{4\,\mathrm{years}}\right)
					\; \left( \dfrac{f_\mathrm{eff}}{80\,\%} \right)
					\; \left( \dfrac{\mathrm{FOV}}{0.833\,\mathrm{deg^2}} \right)
					\; \mathrm{ksec}
					\text{ ,}	
			\end{align}
		where  $t_\mathrm{survey}$ is the survey duration and $f_\mathrm{eff}$ is the observing efficiency, whose expected value is $f_\mathrm{eff}\approx80\,\%$,
		and the eROSITA field-of-view (FOV) is $1.03^\circ$ in diameter (see the SB). 
		The average numbers of background counts to be accumulated  within the PSF HEW in the course of the four-years survey (average exposure time of $\approx2.0\;\mathrm{ksec}$ per point) are $\approx0.8$ and $\approx0.6$ in the soft and hard band.
		For these numbers and for the chosen confidence level, the source detection thresholds are $\approx8$ and $\approx7$ source counts within the PSF HEW.
		The corresponding point-source detection sensitivities are $\left<S_\mathrm{lim}\right>\approx1.1\times10^{-14}$ and $\approx1.8\times10^{-13}\;\mathrm{erg\;s^{-1}\,cm^{-2}}$ in the soft and hard band.
		
		After the first half year of the survey, eROSITA will have scanned the whole sky once.
		At the averaged exposure time of $\approx260\;\mathrm{sec}$, there will be $\approx0.1$ background counts per PSF HEW in each energy band and the point-source detection threshold  will be $\approx4$ counts.
		The point-source detection limits for the half-year survey are $\approx4.8\times10^{-14}$ and $\approx8.0\times10^{-13}\;\mathrm{erg\;s^{-1}\,cm^{-2}}$.

		The main characteristics of the full survey and its first half year are summarized in Table~\ref{tab:Survey}.
		The numbers  are generally consistent with the SB.
		The small differences are related to the differences in the background estimates and the larger PSF size used here for the hard band.
		We also iteratively calculated the resolved fraction of the extragalactic CXB for sensitivities.
				
 		Thus, the eRASS will have on average an $\sim30$ times better sensitivity in the soft band than the previous all-sky survey in this band conducted by ROSAT \citep{Voges1999}.
 		On the other hand, its sensitivity is between one to four orders of magnitude lower than that of the deep but much more narrow (some of them pencil-beam) extragalactic X-ray surveys conducted by \textit{Chandra} and XMM-Newton, such as CDFs, COSMOS, Bootes, Lockman Hole, and ChaMP \citep[see][for a review]{Brandt2005}.

		\subsection{Confusion limit}  \label{ssec:ConL}
		
		For the purpose of this study we assumed that the source confusion becomes important at a source density of one sources per $40$ telescope beams ($=\mathrm{PSF\;HEW}$), which for the angular resolution of eROSITA corresponds to a source density of $\approx460$ and $\approx260\;\mathrm{sources\;deg^{-2}}$ in the soft and hard band.
		With the average of the extragalactic $\log N - \log S$ curves of G08 and \citet[Table~3, ChaMP+CDFs, hereafter K07]{Kim2007}, the corresponding flux levels are $\approx2.3\times10^{-15}$ and $\approx1.2\times10^{-14}\;\mathrm{erg\;s^{-1}\,cm^{-2}}$. 
		In the soft band the confusion limit is achieved at an exposure time of $\approx20\;\mathrm{ksec}$.
		In the hard band, source confusion, achieved at an exposure time of $\approx190\;\mathrm{ksec}$, is not a problem for eRASS.

		
			\begin{figure*}[htp]
				\centering   
				{\includegraphics[width=17.5cm]{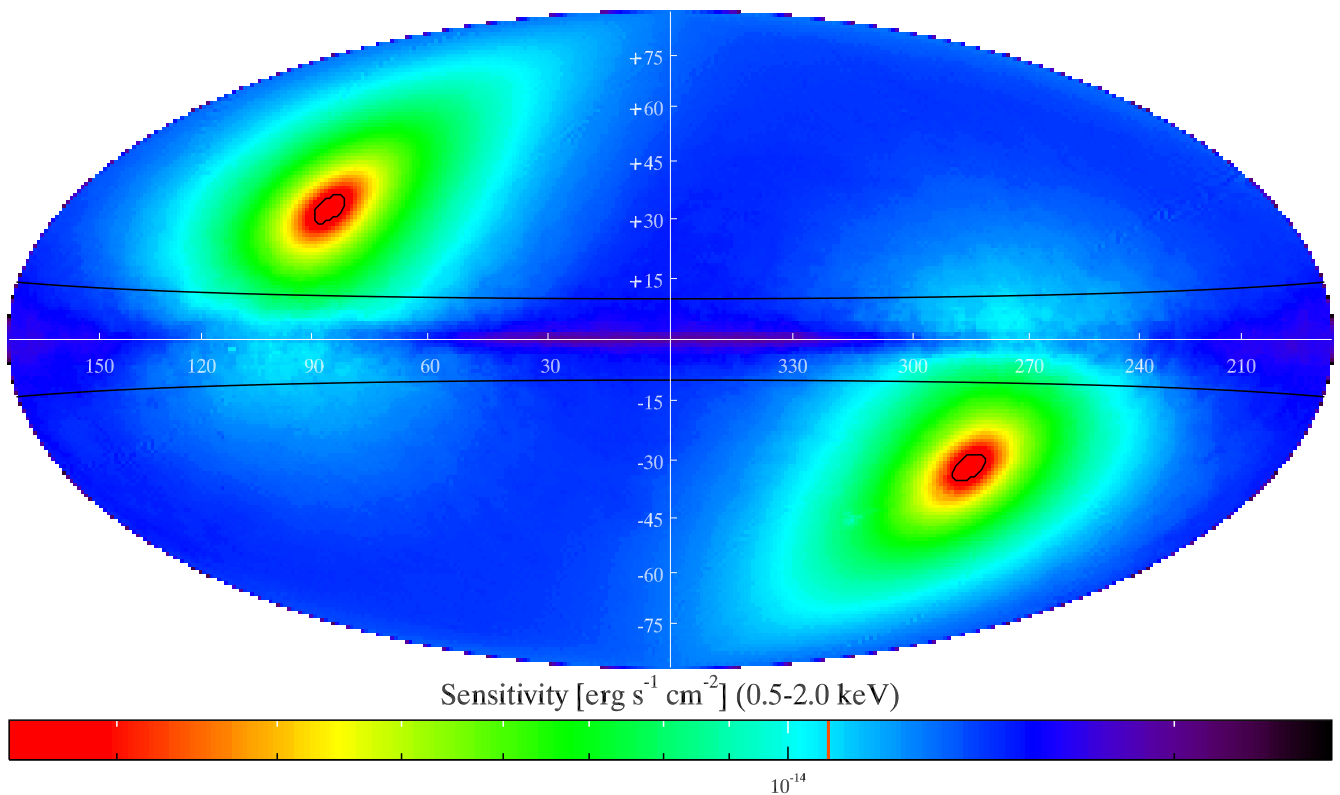}}
				\caption{\label{fig:S_Map}%
					Four-year soft-band sensitivity map of eRASS in Galactic coordinates ($l$, $b$), based on the exposure time map of J.~Robrade (priv.\ comm.) for a continuous Sun-pointing and based on the $N_\mathrm{H}$-map of \citet{Kalberla2005}.
					The two black horizontal curves enclose the Galactic plane ($|b|<10\degr$), which is excluded from our calculation, and the two regions encircled by black curves are our defined ecliptic poles, where the exposure time was set to $20.0\;\mathrm{ksec}$.
					The red vertical line in the horizontal color bar shows the average sensitivity (from Table~\ref{tab:Survey}).
				}
			\end{figure*}
		
		
			\begin{figure}  
				\resizebox{\hsize}{!}{\includegraphics{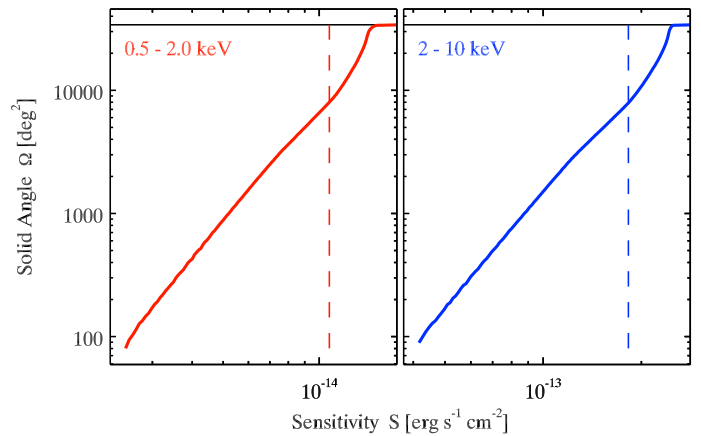}}  
				\caption{\label{fig:Omega}
					The sensitivity -- solid angle curves for the soft (\emph{left panel}) and hard (\emph{right}) bands.
					The vertical dashed lines show the corresponding average sensitivities from the Table~\ref{tab:Survey}.
					The horizontal line shows the solid angle of the extragalactic sky.
				}
			\end{figure}
			
		\subsection{Sensitivity map}  \label{ssec:SensMap}

		Owing to the properties of the scan pattern, the two ecliptic poles of eROSITA will have a significantly higher exposure time than the sky on average.
		This leads to a higher sensitivity at the ecliptic poles.
		The scan strategy of eROSITA is still under discussion and different scenarios are still possible, depending on whether the satellite rotation axis is continuously pointing at the Sun  or moves around it with a slight offset. 
		In the latter scenario, the ecliptic pole regions will occupy a larger  solid angle and will be less overexposed.
				We consider here the extreme case of the continuous Sun-pointing of the scan axis.

		Using the exposure map of the four-year survey \citetext{Robrade, priv.\ comm.}, we defined the two sky regions at the ecliptic poles of eROSITA, where the exposure time (reduced by the observing efficiency) exceeds the confusion limit of $20.0\;\mathrm{ksec}$.
		These two pole regions combined cover a solid angle  of $\approx90\;\mathrm{deg^2}$.
		The point-source detection sensitivity in the soft band  in the pole regions is defined by the confusion limit and is approximately $2.3\times10^{-15}\;\mathrm{erg\;s^{-1}\,cm^{-2}}$, taking into account $\approx50\,\%$ resolved CXB fraction.
		The survey characteristics for the pole regions are summarized in Table~\ref{tab:Survey}.
		In the hard band, the confusion limit is reached at a much longer exposure time of $\approx190\;\mathrm{ksec}$ and is not relevant for the all-sky survey.
		The actual hard-band sensitivity in the pole regions is determined by the particular scan pattern.
		For reference, we list in the Table~\ref{tab:Survey} the sensitivity that can be achieved in the hard band assuming the $20.0\;\mathrm{ksec}$ exposure.


		Outside the poles, the exposure time still varies significantly, with lowest value of $\approx1.6\;\mathrm{ksec}$ achieved in the equatorial regions.
		These variations will lead to variations of the point-source detection sensitivity across the sky.
		To compute a realistic sensitivity map of the survey, we took into account variations of the Galactic absorption across the sky along with the exposure map.
		To this end we used the $N_\mathrm{H}$-map of \citet{Kalberla2005}. 
		We excluded the Galactic plane and only considered the sky at Galactic latitudes $|b|>10\degr$ for the subsequent calculations.
		In computing the exposure map we assumed an observing efficiency of $f_\mathrm{eff}=80\,\%$  and set overexposed regions at the Galactic poles  to $20.0\;\mathrm{ksec}$.
		The solid angle of this extragalactic sky is $\Omega\approx34\,100\;\mathrm{deg^2}$, which corresponds to $\approx83\,\%$ of the total sky.
		For the extragalactic sky the arithmetic mean of the exposure map is $\approx2.1\;\mathrm{ksec}$, which is close to the  average exposure time computed from  Eq.~\eqref{eq:ET}.

		For background calculations, we assumed a constant count rate  for the particle background.
		We assumed that the soft Galactic background is produced in the Local Bubble and therefore is not subject to Galactic absorption,
		whereas the contribution of the extragalactic CXB component was computed with the $N_\mathrm{H}$-map taking into account.
		The resolved extragalactic CXB fraction was fixed at the sky-average value (see Sect.~\ref{ssec:BKG} and Table~\ref{tab:Survey}).
		We ignored intrinsic variations of the Galactic and extragalactic CXB, which are unrelated to Galactic absorption,.
		The amplitude of their variations across the sky can be inferred from the ROSAT diffuse sky maps, as described in Sect.~\ref{ssec:BKG}.
		Background count-rate variations of such amplitude  will result in sensitivity variations of $\sim30\%$.
		A part of these variations is caused by the variations of the Galactic absorption and is included in our calculations through the $N_\mathrm{H}$-map.
	
		With these assumptions we computed sensitivity maps for the two bands; the one for the soft band is shown in Fig.~\ref{fig:S_Map}.
		The sensitivity -- solid angle dependences for both bands are shown in Fig.~\ref{fig:Omega}.

		
			\begin{figure*}[htp]
				\centering
				\includegraphics[width=18cm]{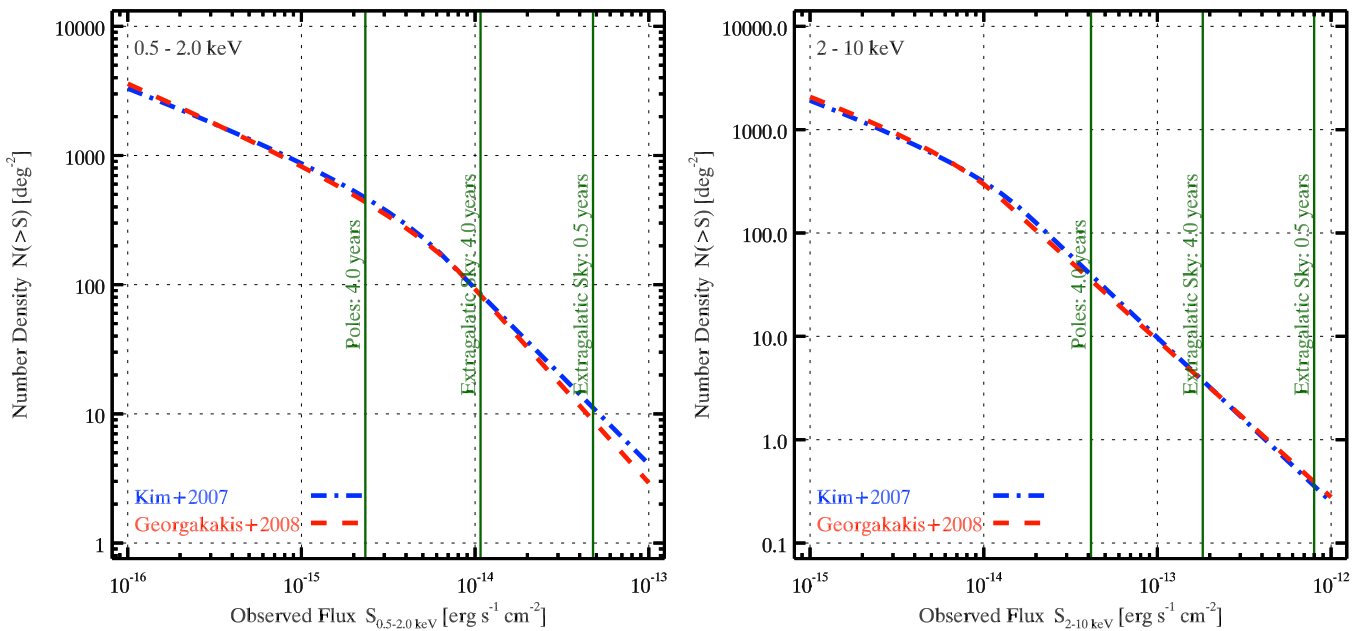}
				\caption{\label{fig:N_S}%
				Cumulative number counts $N(>\!S)$ for observed X-ray point-sources in the soft (\emph{left}) and hard band (\emph{right}).
				The blue dashed-dotted curve is from \citet[Table~3, ChaMP+CDFs]{Kim2007} and the red dashed curve from \citet{Georgakakis2008}.
				The vertical solid lines indicate the point-source detection  sensitivity for different survey durations assuming  $80\,\%$ observing efficiency, and the sensitivity at the 20 ksec exposure time, corresponding to the confusion limit in the soft band (left-most line marked "Poles").
				}
			\end{figure*}		

		\section{AGN number counts}  \label{sec:LogNLogS}
		
		To estimate the source densities and the total numbers of sources detected in different bands we used the source counts results  of K07 and G08, displayed in Fig.~\ref{fig:N_S}.
		For K07, we used the best-fit parameters for the ChaMP+CDFs data from their Table~3 and converted the break flux ($S_\mathrm{b}$) into the hard band from $2.0-8.0\mathrm{\,keV}$ to $2.0-10.0\mathrm{\,keV}$ assuming a power-law spectrum with a photon index of $\Gamma = 1.4$, as was used by K07.
		The best-fit parameters of K07 and G08 agree well (within $2\sigma$).
		The difference in number counts between the two $\log N - \log S$ curves is mostly below $10\,\%$, the strongest deviation is about $20\%$ in the flux range of interest (defined by the characteristic sensitivities, see Table~\ref{tab:Survey}). 
		In the following calculations we used the average of the values given by  the two $\log N - \log S$ curves.		
		
		
		With these curves and our sensitivity map in each energy band, we computed the number density map.	
		The arithmetic mean of this map gives us an average number density of $\approx81\;\mathrm{deg^{-2}}$ and $\approx3.8\;\mathrm{deg^{-2}}$ in the soft and hard band.
		The total numbers of sources detected are $\approx2.7\times10^6$ and $\approx1.3\times10^5$ for the extragalactic sky. These values differ slightly from those in Table~\ref{tab:Survey} because the latter were computed using the average sensitivities of the survey.
		They agree reasonably well with those in the SB.
		About $10\,\%$ of these sources in both bands will be  detected after the first half year of the survey.

		Taking the corresponding sensitivities from Table~\ref{tab:Survey}, we computed a number density of $\approx450\;\mathrm{deg^{-2}}$ and $\approx37\;\mathrm{deg^{-2}}$ in the soft and hard band for the ecliptic poles.
		This translates into $\approx41\,000$ and $\approx3\,400$ detected extragalactic point-sources after four years.
		
 		About $10\,\%$ of the brightest AGN in eRASS will be detected with at least $\approx38$ and $\approx30\;\mathrm{counts}$ per PSF HEW (corresponding flux limits $\approx5.4\times10^{-14}$ and $\approx8.0\times10^{-13}\;\mathrm{erg\;s^{-1}\,cm^{-2}}$) in the soft and hard band.
 		The faintest $10\,\%$ will have approximately $8$ and $7\;\mathrm{counts}$ per PSF HEW ($\approx1.2\times10^{-14}$ and $\approx2.0\times10^{-13}\;\mathrm{erg\;s^{-1}\,cm^{-2}}$).

		We used the sensitivity map to estimate the numbers of AGN expected to be detected in the Galactic plane, $|b|<10\degr$, and obtained $\approx4.6\times10^5$ and $\approx2.2\times10^{4}$ sources.
		This is a somewhat lower number than predicted using the average source density on the extragalactic sky ($\approx6.0\times10^5$ and $\approx2.7\times10^{4}$) because the Galactic absorption in the Galactic plane is on average an order of magnitude higher than for the extragalactic sky.
		The higher Galactic X-ray background, not accounted for in these calculations, will additionally reduce the number of AGN at low Galactic latitudes, and high confusion with Galactic sources will make identifying them more difficult.
		
		Finally, we note that AGN will be the most abundant source in eRASS.
		In addition, eRASS will detect about $\sim10^5$ galaxy clusters \citep{eROSITA}, $\sim2\times10^4$ normal galaxies \citep{Prokopenko2009}, and $\sim4\times10^5$ stars \citep{eROSITA.SB}.

		\section{X-ray luminosity function of AGN} \label{sec:XLF}
		\label{sec:xlf}
		
		With the knowledge of the X-ray luminosity function (XLF) of AGN,
			\begin{align} \label{eq:XLF_gen} 
				\phi(L,z) & = \dfrac{\mathrm{d}\Phi(L,z)}{\mathrm{d}\!\log L} 
				\qquad \text{,}	
			\end{align}
		we are able to compute the redshift and luminosity distributions of the AGN expected to be detected in eRASS.
		The XLF describes the number of AGN ($N$) per unit co-moving volume ($V$) and logarithmic X-ray luminosity ($\log L$) as a function of X-ray luminosity $L$ and redshift $z$.
		It is currently believed that the  luminosity-dependent density evolution (LDDE) model describes the shape of the observed XLF best \citep{Miyaji2000, Ueda2003, Hasinger2005, Silverman2008, Ebrero2009}. 
		For completeness, we summarize it below.
		The LDDE model that parameterizes the AGN XLF has a double power-law
			\begin{align} \label{eq:XLF_LDDE} 
				\phi(L,z) & =	K_0 \; \left[ \left( \dfrac{L}{L_\ast} \right)^{\gamma_1}
						+ \left( \dfrac{L}{L_\ast} \right)^{\gamma_2} \right]^{-1}
						\; e(L,z)
				\qquad \text{,}	
			\end{align}
		with the density evolution factor given by 
			\begin{align} \label{eq:DEF} 
				e(L,z) & =	\left\{
							\begin{array}{l@{\qquad}l}
								(1+z)^{p_1} & z \leq z_\mathrm{c}(L) \\
								\bigl(1+z_\mathrm{c}(L)\bigr)^{p_1} \; \left( \dfrac{1+z}{1+z_\mathrm{c}(L)} \right)^{p_2}  & z > z_\mathrm{c}(L)
							\end{array}
						\right.
				\qquad \text{,}	
			\end{align}
		where the cutoff redshift is given by
			\begin{align} \label{eq:z_c} 
				z_\mathrm{c}(L) & =	\left\{
							\begin{array}{l@{\qquad}l}
								z_\mathrm{c,0} \; \left( \dfrac{L}{L_\alpha} \right)^\alpha & L \leq L_\alpha \\
								z_\mathrm{c,0}  & L > L_\alpha
							\end{array}
						\right.
				\qquad \text{,}	
			\end{align}
		This LDDE model has nine parameters. 
 		\citet{Miyaji2000} defined the density evolution factor (Eqs.~\ref{eq:DEF} and \ref{eq:z_c}) for the soft-band XLF in a slightly different way, but the concept remains the same.
 		 \citet{Hasinger2005} used the luminosity-dependent density evolution indices ($p_1$ and $p_2$)
		\begin{align} 
 			p_1(L) & = p_{1_{44}} + \beta_1 \; (\log L - 44.0) \qquad \text{} \label{eq:p1} 	\\
 			p_2(L) & = p_{2_{44}} + \beta_2 \; (\log L - 44.0) \qquad \text{} \label{eq:p2}
			\qquad \text{,}	
		\end{align}
 		and therefore the number of parameters increases with the two additional parameters ($\beta_1$ and $\beta_2$) to eleven.
		
		As our default XLF models we used the LDDE model of \citet[][Table~5, hereafter H05]{Hasinger2005} for the soft band and of \citet[hereafter A10]{Aird2010} for the hard band.
		For the hard-band XLF we used the best-fit model from A10, the "color preselected sample" (their Table~4), which is expected to provide a more accurate description of the XLF at higher redshifts.
		The parameters of the chosen XLF models are summarized in Table~\ref{tab:LDDE_BestFit}.

		
			\begin{table}
				\caption{\label{tab:LDDE_BestFit} Parameters of the LDDE model used to compute the luminosity and redshift distributions of the detected AGN.}  
				\centering
				\begin{tabular}{c c c}
					\hline\hline
					Energy band [keV]       & $0.5-2.0$ & $2-10$ \\
					XLF & H05 (Table~5)  & A10 (Table~4) \\
					\hline	\\ [-1ex] 
					$K_{44}\;/\;K_0$ \tablefootmark{(a)}	& $2.62 \pm 0.16$} \tablefootmark{(b)	& $8.32 \pm 1.15$ \\
					$\log L_\ast$ \tablefootmark{(c)}		& $43.94 \pm 0.11$ 		& $44.42 \pm 0.04$\\
					$\gamma_1$		& $0.87 \pm 0.10$	& $ 0.77 \pm 0.01$ \\
					$\gamma_2$		& $2.57 \pm 0.16$	& $ 2.80 \pm 0.12$ \\
					$p_{1_{44}}\;/\;p_1$	& $ 4.7 \pm 0.3$ \tablefootmark{(d)}	& $4.64 \pm 0.24$ \\
					$p_{2_{44}}\;/\;p_2$	& $-1.5 \pm 0.7$ \tablefootmark{(d)}	& $-1.69 \pm 0.12$ \\
					$z_\mathrm{c,44}\;/\;z_\mathrm{c,0}$	& $1.42 \pm 0.11$ \tablefootmark{(e)}	& $ 1.27 \pm 0.07$\\
					$\log L_\alpha$ \tablefootmark{(c)} 		& $44.67 \;\mathrm{(fixed)}$ 	& $44.70 \pm 0.12$\\
					$\alpha$		& $0.21 \pm 0.04$	& $ 0.11 \pm 0.01$ \\
					$\beta_1$		& $ 0.7 \pm 0.3$	& -- \\
					$\beta_2$		& $ 0.6 \pm 0.8$	& -- \\ [+1ex] 
					\hline
				\end{tabular}
				\tablefoot{
					\tablefoottext{a}{In units of $10^{-7}\,\mathrm{Mpc^{-3}}$.}
					\tablefoottext{b}{
						$ K_0 =  K_{44} \; [ (10^{44.0}/L_\ast)^{\gamma_1} + (10^{44.0}/L_\ast)^{\gamma_2} ] \approx 6.69$
						.}
					\tablefoottext{c}{$\mathrm{erg\;s^{-1}}$.}
					\tablefoottext{d}{$p_1$ and $p_2$ are computed from Eqs.~\eqref{eq:p1} and \eqref{eq:p2}.}
					\tablefoottext{e}{ $z_\mathrm{c,0} = z_\mathrm{c,44} \; 10^{\alpha\,(\log L_\alpha - 44.0)} \approx 1.96$.}
				}
			\end{table}


		
			\begin{figure*} 
				\centering   
				{\includegraphics[width=17.5cm]{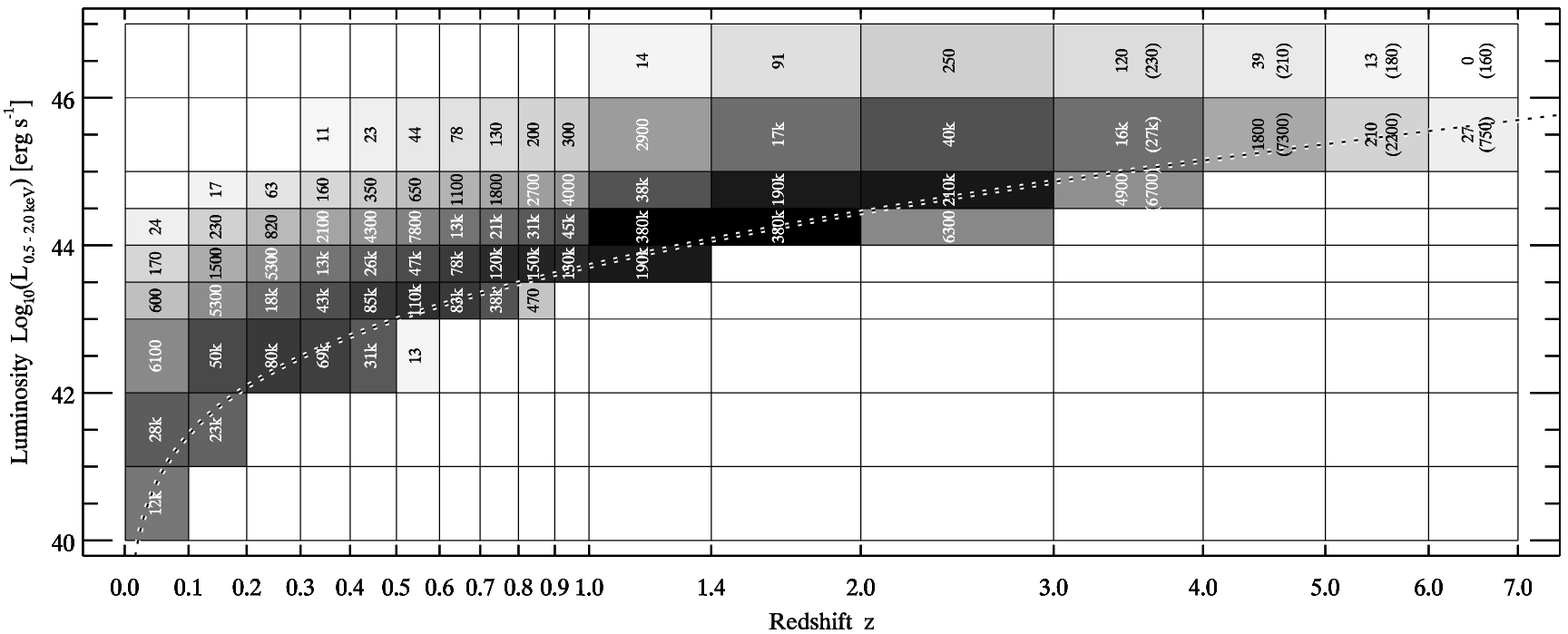}}
				{\includegraphics[width=17.5cm]{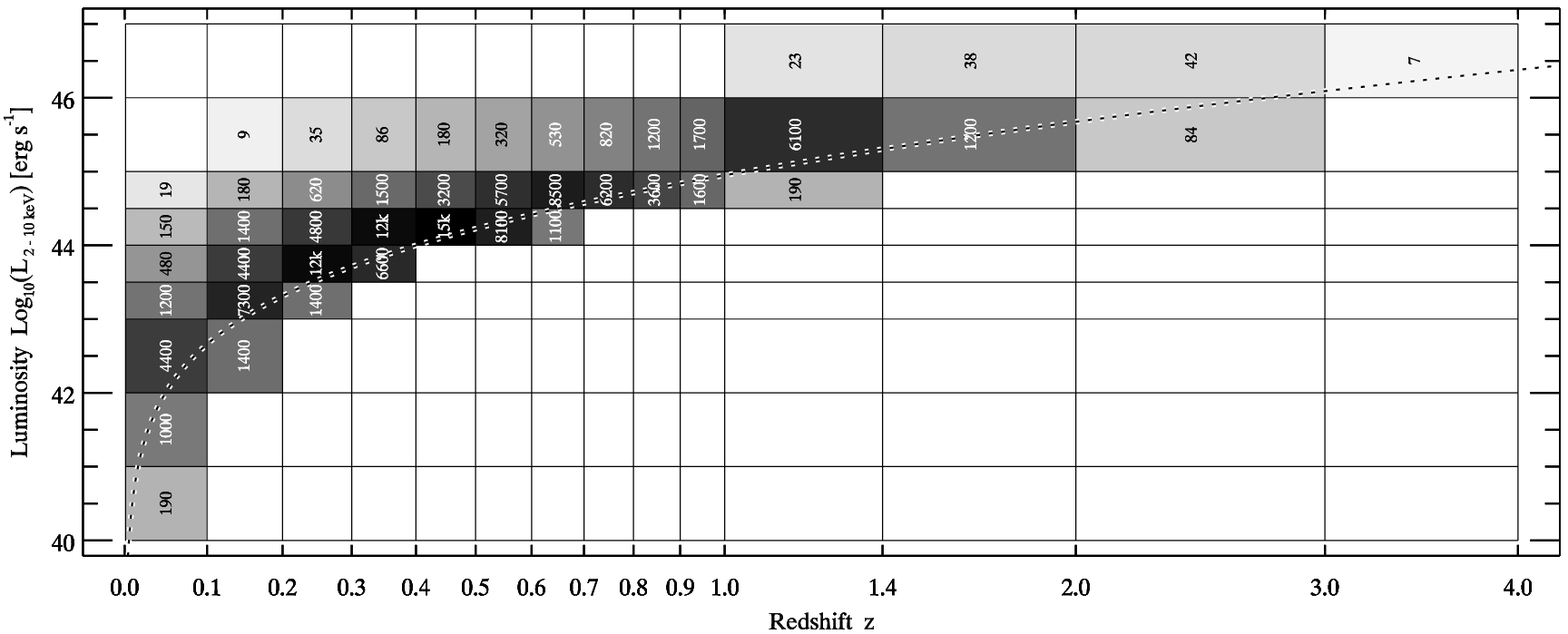}}
				\caption{\label{fig:N_bin}%
					Number of AGN in different redshift and luminosity bins expected to be detected in the course of the four-year survey in the soft (\emph{top})  and  hard (\emph{bottom}) bands.
					White empty bins with no number correspond to zero sources.
					The dotted line corresponds to the detection limit of eRASS.
					In the soft-band plot, the numbers in brackets are for the XLF without the exponential redshift cutoff, they are given only if the difference exceeds $10\,\%$.
				}
			\end{figure*}

		Based on the XMM-Newton data, \citet{Brusa2009} demonstrated  that the soft-band XLF of H05 overpredicts  the numbers of high-redshift objects, $z>3$, detected in the COSMOS survey.
		These authors proposed to introduce an exponential redshift cutoff of the XLF for $z>2.7$,
		\begin{align} \label{eq:ExpDec} 
			\phi & = \phi_\mathrm{H05}(z\!=\!2.7)\, \times 10^{0.43\,(2.7 - z)}, \qquad z>2.7
			\qquad \text{,}	
		\end{align}
		and showed that with this modification, the observed number counts of high-redshift AGN are reproduced much better.
		This result  was also confirmed by \citet{Civano2011}, who used additional \textit{Chandra} data on the same field and analyzed an  $\approx50\,\%$ larger AGN sample than \citet{Brusa2009},
		and by \citet{Hiroi2012}, who analyzed $30$ high-redshift ($z>3$) AGN in the Subaru/XMM-Newton Deep Survey field.
		Introducing the redshift cutoff results in an insignificant decrease of $\sim1\,\%$ in the total number of AGN above the eRASS sensitivity limit. 
		However, it has a strong effect on the numbers of high-redshift objects, which we discuss in Sect.~\ref{ssec:HighzAGN}.
		For our default XLF in the soft band, we  included the high-redshift cutoff described by Eq.~\eqref{eq:ExpDec}, but additionally show results without cutoff.
		
		As a consistency check, we computed the $\log N - \log S$ distributions based on the chosen XLF models and compared them with the results of the source counts by K07 and G08. 
 		The $\log N - \log S$ can be computed by integrating the XLF over luminosity $L$ and redshift $z$: 
		\begin{align} \label{eq:NS_gen} 
			N(>\!S)	& = \int\limits_0^{z_\mathrm{max}} \dfrac{\mathrm{d}V(z)}{\mathrm{d}z}
					\int\limits_{\log L_\mathrm{min}(S,z)}^{\log L_\mathrm{max}} \phi(\log L,z) \,\mathrm{d}\log L \; \mathrm{d}z
			\qquad \text{.}	
		\end{align}
		Here, $\tfrac{\mathrm{d}V(z)}{\mathrm{d}z}$ $[\mathrm{Mpc^3\,sr^{-1}}]$ is the co-moving volume element per redshift and solid angle%
			\footnote{
				The solid angle is converted from steradian to square degrees ($\pi^2\;\mathrm{sr}=180^2\;\mathrm{deg^2}$).
			}
		and $L_\mathrm{min}(S,z) = 4\pi\,S\,d_\mathrm{L}^2(z)$, where $d_\mathrm{L}(z)$ is the luminosity distance \citep[e.g.][]{Hogg1999}.
		K-correction was applied, assuming a power-law spectra with the photon index $\Gamma = 1.9$ and no absorption.
		The same photon index was used to convert the XLFs to the energy bands used in this paper, if the former was determined for a different energy band.
		It is worth to mention that deep X-ray surveys do not show any evidence of a redshift-dependent photon index \citep{Brandt2005}. 
	
		In Eq.~\eqref{eq:NS_gen} as well as in the calculations described in the next Sections, we integrated the XLF model in the luminosity range of $40 \leq \log L[\mathrm{erg\;s^{-1}}] \leq 48$ and in the redshift range of $0\leq z\leq7$.
		Decrease of the $L_\mathrm{min}$ in the luminosity integration or increase of the upper limit for the redshift integration, has no significant effect  on the number counts $N(>\!S)$ in our flux range of interest.
		We note that all experimental XLF determinations are based on AGN samples, that cover a smaller luminosity range, typically $L\geq10^{42}\;\mathrm{erg\, s^{-1}}$, and a smaller redshift range ($z_\mathrm{max}\approx3-5$).
		Hence, our calculations involve some extrapolation of the measured  XLFs to lower luminosities and higher redshift.
		The uncertainties introduced  by this extrapolation are generally small, with a few exceptions that are discussed below.
		
		Using the XLF of H05, we predict a somewhat smaller number of AGN for the soft-band counts than that observed by K07 and G08, with a strongest deviation of about $30-50\,\%$ for the $\log N - \log S$ curve in our flux range of interest. 
		Part of this discrepancy arises because H05 selected only Type 1 AGN, and part may be caused by cosmic variance.
		It is beyond the scope of this work to investigate the origin of this difference in detail, therefore we renormalized the soft-band XLF of H05 upward with a factor of $\approx1.35$ to match the source counts of K07 and G08  in the flux range of interest.
		The hard-band $\log N - \log S$ obtained using the XLF of A10 agrees well with the observed source counts, with a strongest deviation of about $3-11\,\%$ in the flux range of interest. 

	
			\begin{figure*}
				\centering
 				\includegraphics[width=16.5cm]{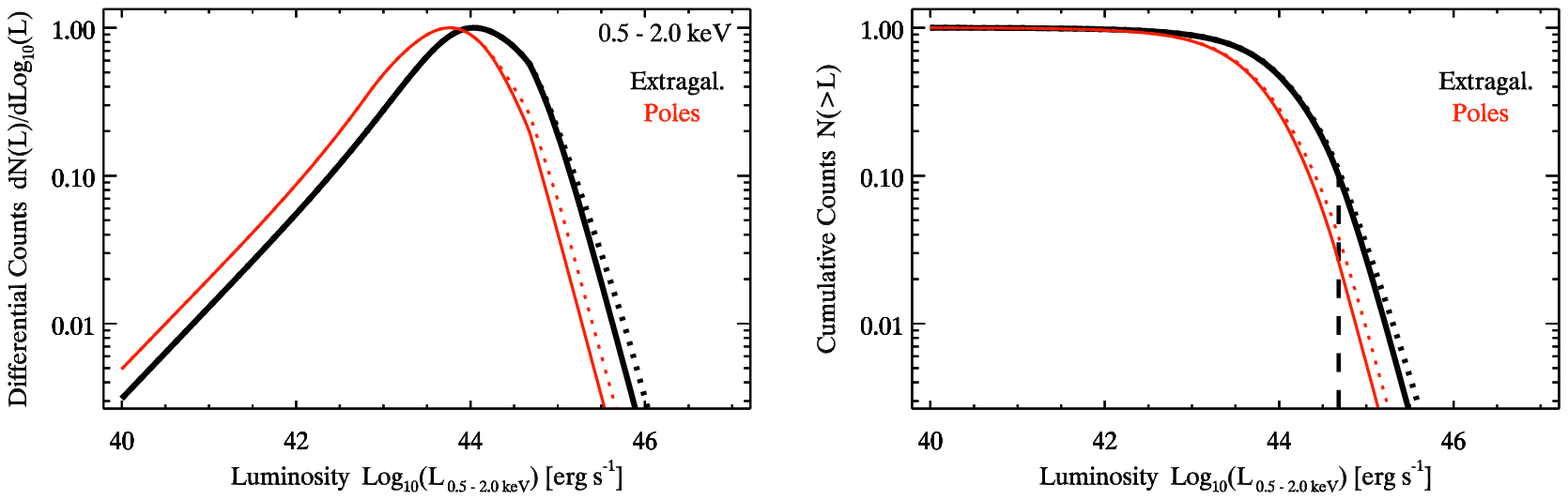}
				\includegraphics[width=16.5cm]{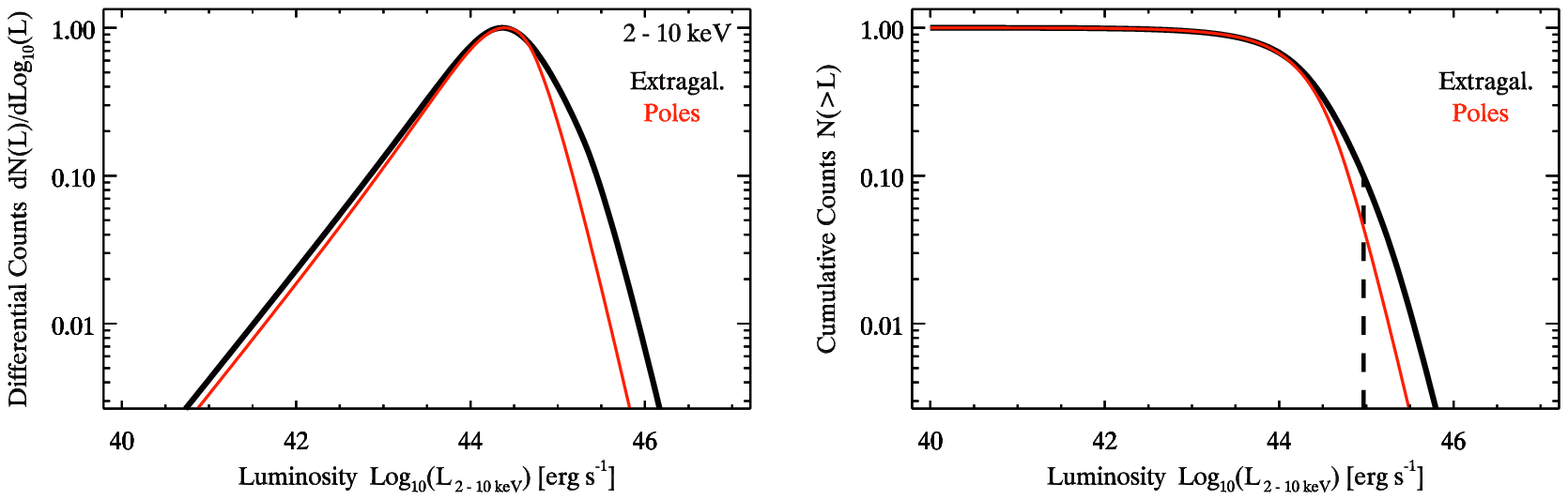}
				\caption{\label{fig:N_L}%
					Differential (\emph{left}) and cumulative (\emph{right}) luminosity distribution in the soft (\emph{top}) and the hard (\emph{bottom}) band for the four-year eRASS AGN sample in the extragalactic sky (\emph{black}) and at the ecliptic poles only (\emph{red}).
					The distributions are normalized to unity to facilitate comparison of the shapes.
					The dotted curves in the top panels were computed without the high-redshift cutoff in the soft-band XLF (see Sect.~\ref{sec:xlf}).
					The dashed black vertical lines in the right panels show the luminosity corresponding to the $10\,\%$ fraction of sources. 
					}
			\end{figure*}

		\section{ Luminosity and redshift distribution of detected AGN} \label{sec:NL_Nz}
		
		
		With the knowledge of the XLF (Sect.~\ref{sec:XLF}) we can compute luminosity and redshift distributions of detected AGN as follows:
			\begin{align} 
				\dfrac{\mathrm{d}N(L)}{\mathrm{d}\log L} & =
					\int\limits_0^{z_\mathrm{max}(S,L)}\, \phi(\log L,z)  \; \dfrac{\mathrm{d}V(z)}{\mathrm{d}z} \; \mathrm{d}z
					\label{eq:dNdL}  \\
				\dfrac{\mathrm{d}N(z)}{\mathrm{d}z} & =
					\dfrac{\mathrm{d}V(z)}{\mathrm{d}z} \; \int\limits_{\log L_\mathrm{min}(S,z)}^{\log L_\mathrm{max}} \phi(\log L,z) \,\mathrm{d}\log L
					\label{eq:dNdz}
				\qquad \text{,}	
			\end{align}
		where $z_\mathrm{max}$ is defined by the relation $d_\mathrm{L}(z_\mathrm{max})=\sqrt{L/(4\pi\,S)}$, where  $d_\mathrm{L}(z_\mathrm{max})$ is the luminosity distance at the redshift  $z_\mathrm{max}$.
		For the other quantities and the K-correction see the explanation after Eq.~\eqref{eq:NS_gen}.
		The corresponding cumulative distributions are
			\begin{align} 
				N(>\!L)	& = \int\limits_L^{L_\mathrm{max}} \mathrm{d}N(L')
					\label{eq:N_L}  \\
	%
				N(>\!z) & = \int\limits_z^{z_\mathrm{max}} \mathrm{d}N(z')
					\label{eq:N_z} 
			\end{align}
	
		The number of AGN detected in the eRASS as a function of luminosity and redshift is summarized in Fig.~\ref{fig:N_bin} and is discussed in more detail in the next two subsections.
		In computing these distributions we took into account the sensitivity map (Sect.~\ref{ssec:SensMap}) of the eRASS via the sensitivity -- solid angle distribution shown in Fig.~\ref{fig:Omega}.
		For the overexposed areas at the ecliptic poles we used the sensitivity quoted in Table~\ref{tab:Survey}. 
		The properties of the brightest and faintest $10\,\%$ we computed using the flux limits from Sect.~\ref{sec:LogNLogS}.

	
			\begin{figure*}
				\centering
				\includegraphics[width=16.5cm]{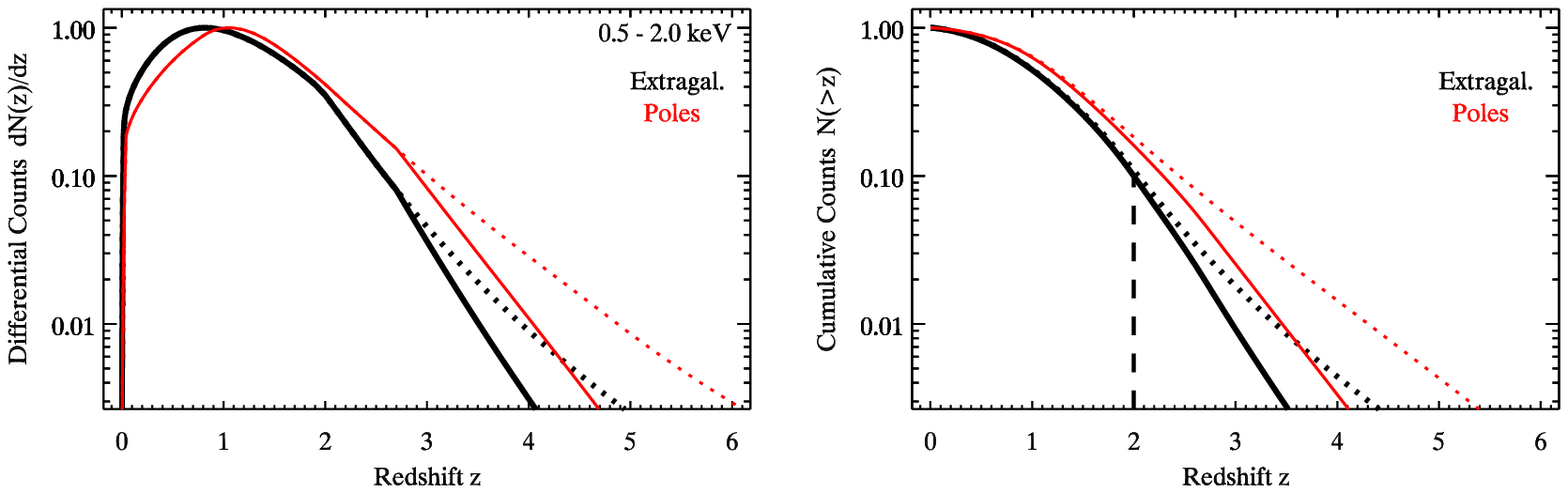}
				\includegraphics[width=16.5cm]{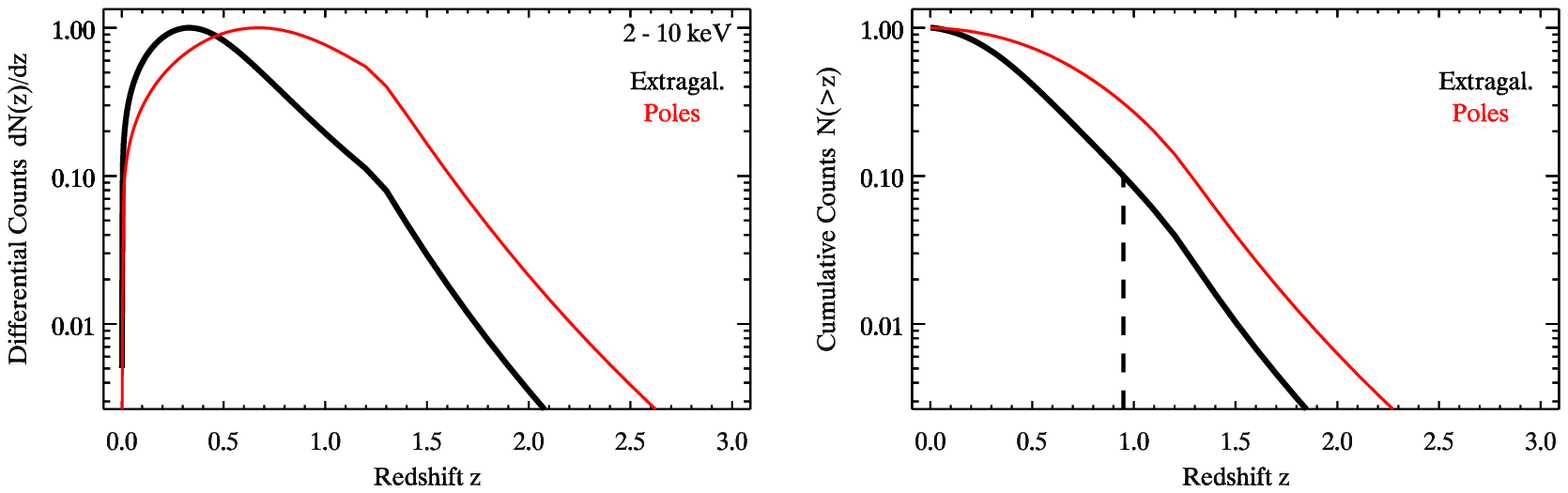}
				\caption{\label{fig:N_z}%
					Same as Fig.~\ref{fig:N_L}, but for the redshift distribution.
					}
			\end{figure*}

	
			\begin{figure*}
				\centering
				\includegraphics[width=16.5cm]{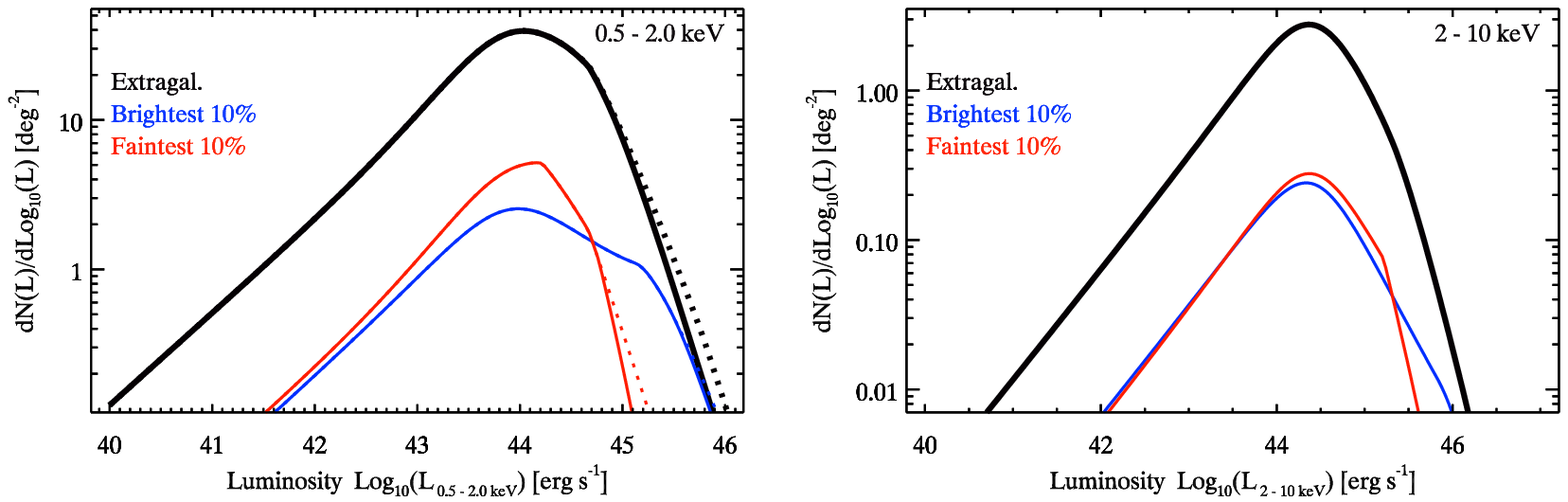}\vspace{0.5cm}
				\includegraphics[width=16.5cm]{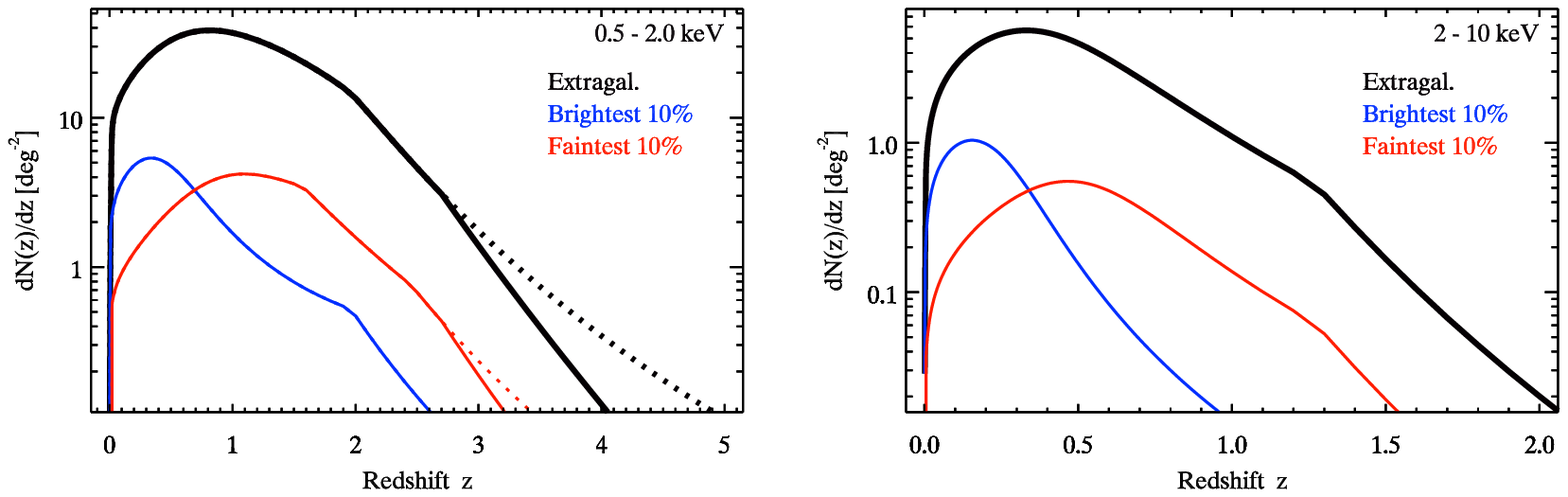}
				\caption{\label{fig:N_fb}%
					Differential luminosity (\emph{top}) and redshift (\emph{bottom}) distributions in the soft (\emph{left}) and hard (\emph{right}) bands for the entire extragalactic AGN sample after fours years (\emph{black}) and for the brightest $10\,\%$ (\emph{blue}) and faintest $10\,\%$ (\emph{red}).
					The black curves are same as in Figs.~\ref{fig:N_L} and \ref{fig:N_z},  but without renormalization to unity.
					}
			\end{figure*}

		\subsection{Luminosity distribution} \label{ssec:LumDis}
		
		Luminosity distributions of detected AGN are shown in Fig.~\ref{fig:N_L}.
		In the soft band they peak at $\sim10^{44}\mathrm{\,erg\;s^{-1}}$, with only a small difference between the extragalactic sky sample ($\approx10^{44.0}\mathrm{\,erg\;s^{-1}}$) and the ecliptic poles ($\approx10^{43.8}\mathrm{\,erg\;s^{-1}}$).
		The peak in the hard band occurs at $\approx10^{44.4}\mathrm{\,erg\;s^{-1}}$.
		The median values differ by less than $1\,\%$ from the corresponding peak values.
		Comparing these with the values of $L_\ast$ from the Table~\ref{tab:LDDE_BestFit},
		the location of the peak is defined by the $L_\ast$ luminosity and does not strongly depend on the survey sensitivity.
		A change of the latter by two orders of magnitude changes the position of the peak only by $\sim0.5\;\mathrm{dex}$. 
		Hence, our predictions for the luminosity distribution are very robust against moderate changes of the survey sensitivity.
		From the top panel of Fig.~\ref{fig:N_L} one can see that the luminosity distribution in the soft band changes only marginally at high luminosity ($\ga10^{44}\mathrm{\,erg\;s^{-1}}$) if we exclude the high-redshift cutoff of the XLF.
		
		From the cumulative luminosity distributions of the extragalactic sky (right panels of Fig.~\ref{fig:N_L}) about $10\,\%$ (vertical dashed lines) of the detected AGN will have luminosities higher than $\sim10^{45}\mathrm{\,erg\;s^{-1}}$.
		This large  sample of luminous AGN ($\sim3\times10^5$ in full redshift range) will improve the constraints on the high-luminosity end of the XLF.
		For comparison, the AGN sample of H05 had about $\sim100$ AGN with a luminosity higher than $\sim10^{45}\mathrm{\,erg\;s^{-1}}$.
		
		In the top panel of Fig.~\ref{fig:N_fb} the luminosity distribution of the brightest $10\,\%$ (those with the highest flux, blue curve) and the faintest $10\,\%$ (lowest flux, red) do not differ much from each other and from the distribution of the total sample (black).
		We note that the luminosity distribution of objects detected in the half-year survey is well represented by that of the brightest $10\,\%$ of the sources.

		
			\begin{table}
				\caption{Peak and median values of the redshift distribution of eRASS AGN}
				\label{tab:N_z}
				\renewcommand{\arraystretch}{1.3} 
				\centering 
				\begin{tabular}{l c c | c c} 
					\hline\hline 	
								& \multicolumn{2}{c}{$0.5-2.0\mathrm{\,keV}$}	& \multicolumn{2}{c}{$2-10\mathrm{\,keV}$}	\\
								& Peak		& Median		& Peak		& Median	\\
					\hline  
					4.0 years extragalactic	 & $0.8$	& $1.0$		& $0.3$	& $0.4$	\\
					4.0 years ecliptic poles & $1.0$	& $1.2$		& $0.7$	& $0.7$	\\
					\hline
					Brightest $10\,\%$ 	& $0.3$	& $0.5$		& $0.2$	& $0.2$	\\
					Faintest $10\,\%$ 	& $1.1$	& $1.2$		& $0.5$	& $0.5$	\\ 
					\hline 
					0.5 years extragalactic	& $0.4$	& $0.6$		& $0.2$	& $0.2$	\\
					\hline 
				\end{tabular}
			\end{table}

		\subsection{Redshift distribution} \label{ssec:RedDis}
		
		Unlike the luminosity distributions, redshift distributions of a flux-limited sample are strongly dependent on the limiting flux  (Fig.~\ref{fig:N_z}).
		Correspondingly, the redshift distributions for the extragalactic sky sample and for the poles peak at different redshift, the difference being larger for the hard band.
		The same is true for the median values, which are listed together with the peak values in Table~\ref{tab:N_z}.
		The median and peak values in the soft band do not change significantly when we exclude the exponential high-redshift cutoff from our calculations.
		However, the redshift distribution of AGN at high-redshift does change significantly, which we can see clearly in the top panel of Fig.~\ref{fig:N_z}.
	 	This is discussed in more detail in Sect.~\ref{ssec:HighzAGN}.
		
		The differential distributions show several breaks that are caused by the derivative discontinuities of the LDDE model.
		Another break at the low redshift, $z\sim0$, appears when the low integration limit in Eq.~\eqref{eq:dNdz}, $\log L_\mathrm{min}(S,z)$, becomes equal to the low limit of the interval where the XLF is defined ($L=10^{40.0}\;\mathrm{erg\;s^{-1}}$, Sect.~\ref{sec:XLF}).
		These features are not physical and reflect the deficiencies of the functional form used in the LDDE model.
		However, these deficiencies of the XLF model do not compromise the overall shapes of the redshift (and luminosity) distributions derived in this paper, as long as the overall shape of the AGN X-ray luminosity function is adequately represented by the LDDE model.

		In accord with the note made at the beginning of this Section, redshift distributions of the brightest and faintest $10\,\%$ of the AGN (Fig.~\ref{fig:N_fb})  peak at significantly different redshifts than the overall distributions (black).
		Similar to luminosity distributions, the redshift distributions of the objects detected during the half-year survey are similar to the distributions of the brightest $10\,\%$.
		From the cumulative distributions (right panels in Fig.~\ref{fig:N_z}.) we conclude  that in the soft band, $\approx50\,\%$ of objects in the eRASS sample will be located at $z>1$, whereas $\approx10\,\%$ will be located at $z>2$.
		
		
			\begin{figure} 
  	 			\resizebox{\hsize}{!}{\includegraphics{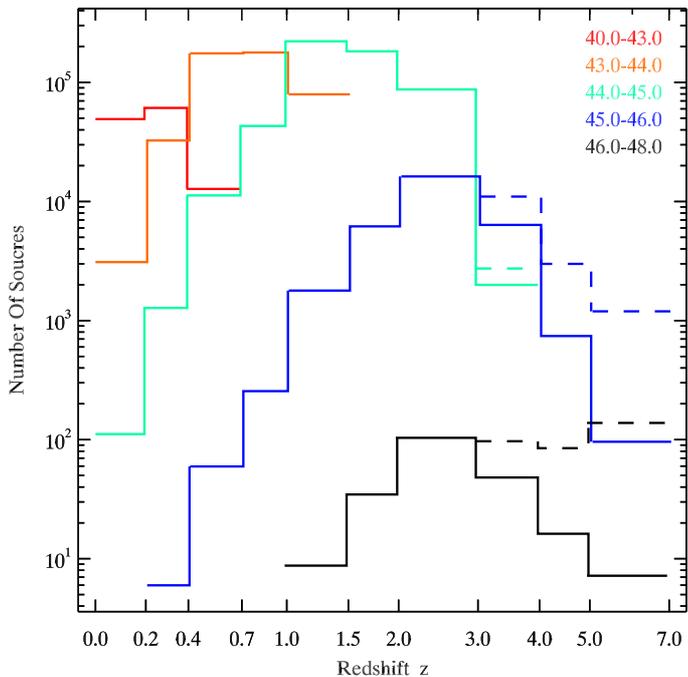}}  
				\caption{\label{fig:XLF}%
					Number of eRASS AGN in the soft band as a function of the redshift for different luminosity groups in a sky area similar to that covered by SDSS ($14\,000\;\mathrm{deg^2}$).
					The solid and dashed histograms show predictions based on our default soft-band XLF with and without exponential high-redshift cutoff (Sect.~\ref{sec:XLF}), respectively.
					}
			\end{figure}
		
		To  illustrate the potential of the eRASS AGN sample in the limited sky areas, we show in Fig.~\ref{fig:XLF} the number of objects per redshift bin as a function of redshift for several luminosity groups.
		For this calculation we chose a sky area of $14\,000\;\mathrm{deg^2}$, similar to the area of the Sloan Digital Sky Survey (SDSS), and considered relatively broad redshift bins, consistent with the expected accuracy of photometric redshifts based on the multiband photometry \citep{Salvato2011}.
		It is obvious from Fig.~\ref{fig:XLF} that even coarse redshift information over relatively limited areas of sky is capable of delivering unprecedented samples of AGN, covering the most luminous AGN ($>10^{45}\mathrm{\,erg\;s^{-1}}$) with an unmatched statistical significance.
				
		
			\begin{figure*} 
				\centering
				\includegraphics[width=16.5cm]{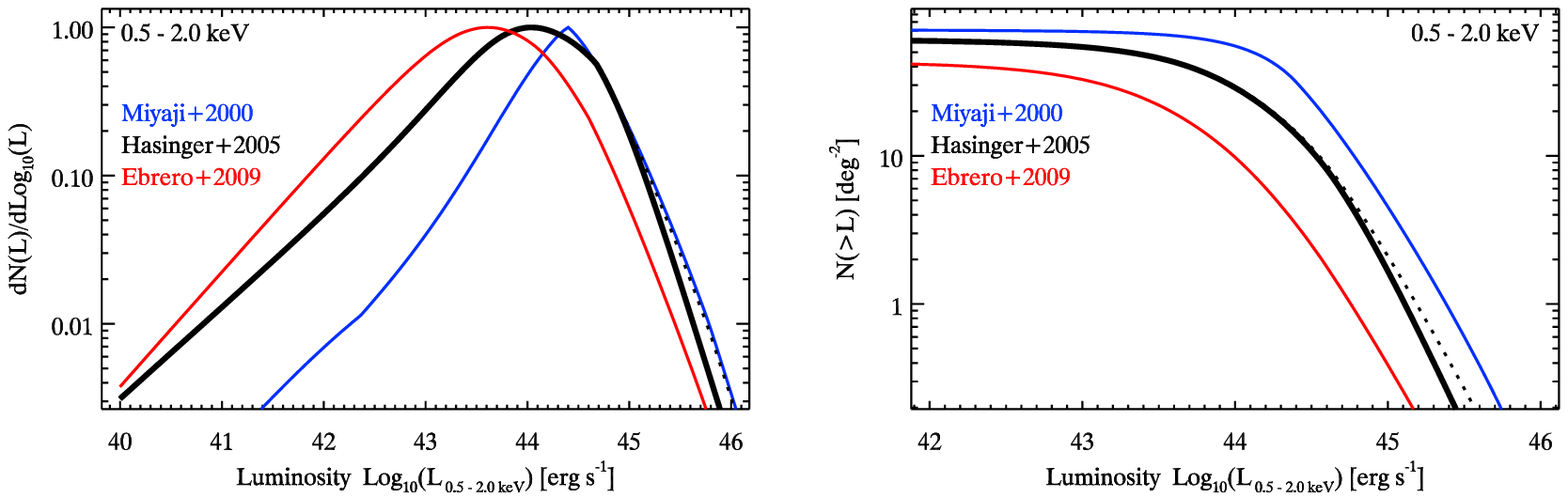}
				\includegraphics[width=16.5cm]{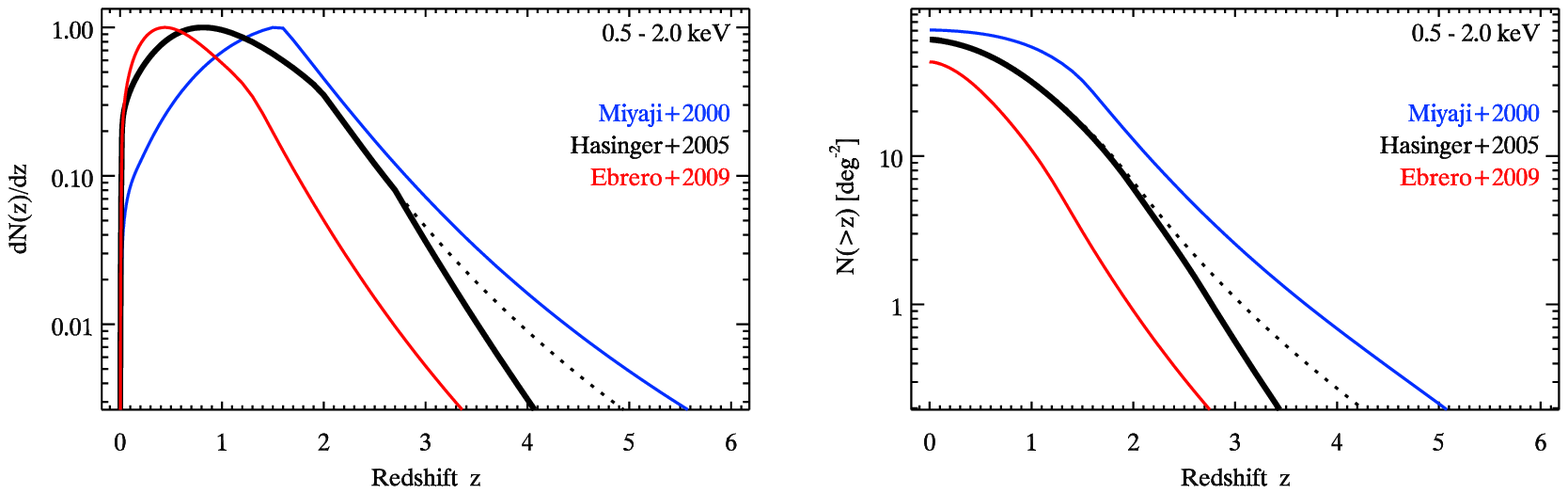}
				\caption{\label{fig:N_Lz}%
					Differential (\emph{left}) and cumulative (\emph{right}) luminosity (\emph{top}) and redshift (\emph{bottom}) distributions of the soft-band extragalactic sky sample computed using different XLFs.
					The thick solid black curves show predictions based on the default XLF model.
					Also shown are predictions for the XLF model of \citet[][Table~3]{Miyaji2000}, \citet[][Table~5, without the exponential redshift cutoff, dotted curve]{Hasinger2005}, and \citet{Ebrero2009}.
					To facilitate the comparison of shapes, differential distributions in left panels are normalized to unity.
					}
			\end{figure*}

		\subsection{Uncertainties}  \label{ssec:RedUnc}
		
		Obviously, the accuracy of our predictions depends on the accuracy of the AGN XLF.
		This is limited by the  moderate numbers of objects used for their construction, typically about $1\,000$.
		Although the XLFs obtained by different authors are broadly consistent with each other, there is still a considerable spread between different models.
		Correspondingly, using XLF models obtained by different authors, we obtained somewhat varying luminosity and redshift distributions of eRASS AGN.
		
		To illustrate the range of uncertainties, we calculated the luminosity and redshift distributions for the soft-band extragalactic sky sample using several different XLF models available in the literature.
		Along with our default soft-band XLF, we used the XLF of H05 without the exponential redshift cutoff and the XLF models of \citet[][Table~3]{Miyaji2000} and \citet{Ebrero2009}.
		These XLF models are based on (overlapping) samples, each containing about $1\,000$ objects in total.
		Because the samples partly overlap, the models are not entirely independent.
		The resulting luminosity and redshift distributions of the eRASS extragalactic sky sample are shown in Fig.~\ref{fig:N_Lz}.
		To facilitate the comparison of the shapes, the differential distributions are normalized to unity,
		whereas the cumulative distributions are shown with their original normalization.

		It can be seen from Fig.~\ref{fig:N_Lz} that different XLFs predict considerably  different luminosity and redshift distributions for eRASS~AGN.
		Although the detailed shapes may be not critically important for the purpose of this paper, median redshifts and luminosities are important characteristics of the eRASS~AGN sample.
		These parameters differ somewhat significantly for the three XLFs, with \citet{Miyaji2000} and  \citet{Ebrero2009} presenting two extremes and our default XLF of \citet{Hasinger2005} located in the middle.
		It is worth mentioning that the primary reason for selecting the XLF of \citet{Hasinger2005} as our default one was the fact that it best reproduced the observed $\log N - \log S$ distributions (Sect.~\ref{sec:XLF}). 
		Note also that the total numbers of eRASS~AGN are predicted sufficiently accurately from the observed $\log N - \log S$ distributions and therefore are not of significant concern.

		In interpreting the plots in Fig.~\ref{fig:N_Lz} one should keep in mind, that different authors applied different  selection criteria in building their samples and used  slightly  different versions of the  LDDE model.
		In addition, \citet{Ebrero2009} applied a correction for absorption, which the others did not.
		These differences explain, in particular, the difference in the total AGN surface density predicted by different models.
		They also explain, at least in part, the considerably large discrepancy in the shapes of the predicted luminosity and redshift distributions.
		Another part of the difference is probably related to statistical uncertainties in the XLF parameters.
		Although a detailed comparison of XLFs produced in different studies is beyond the scope of this paper, it would be interesting to see to which extent the discrepancy can be explained by statistical uncertainties.
		However, the LDDE model is a multiparameter model with a complex correlation pattern between parameters.
		Therefore  a proper error analysis would require knowledge of the covariance matrixes, which are not available anymore (T.~Miyaji, priv.\ comm.).
		An attempt to treat the XLF parameter errors as independent resulted in unreasonably large uncertainties in the predicted distributions for the eRASS sources.
		On the other hand, the moderate size of the statistical errors in the observed XLF data in the redshift and luminosity range relevant to the bulk of eRASS~AGN ($L\sim10^{44}\mathrm{\,erg\;s^{-1}}$, $z\sim1$) suggests that the behavior of the derived distributions near their peaks is probably not significantly affected by the propagation of statistical errors, therefore the differences seen in Fig.~\ref{fig:N_Lz} probably reflect genuine differences in XLFs.

		Another important factor to be taken into consideration is cosmic variance.
		Because the AGN XLF determinations rely on the survey that covers only a small fraction of the sky, $\la10^{-4}-10^{-3}$ at most, they are subject to cosmic variance.
		The amplitude of this uncertainty is probably in the $\sim10\,\%$ range \citep{Aird2010}.
		Obviously, the eRASS sample will provide means for studying this effect in full detail.
		
		We emphasize that we did not consider any separation between different types of AGN.
		H05 only considered type 1 AGN for their XLF model.
		If we take into accout the expected small fraction
		($\sim10\,\%$ \footnote{\url{http://www.bo.astro.it/~gilli/counts.html}},
		see also \citealt{eROSITA.SB}) of type 2 AGN (intrinsic $N_\mathrm{H}> 10^{21}\;\mathrm{cm^{-2}}$)
		and the fairly similar XLF of both types \citep[e.g.][]{Burlon2011},
		we expect that the introduced uncertainties will be relatively small.
		
		Finally, we did not take into account the Eddington bias, neither did we consider the details of the source detection and background subtraction algorithms, which will affect to some extent the numbers of detected sources and their $\log N - \log S$ distributions at the faint end.
		They will also affect the completeness characteristics of the eRASS AGN sample, which will have to be accounted for in constructing XLFs.
		These are typical properties of flux-limited surveys, especially those conducted in the photon-counting regime, in the limit of small numbers of counts, where the character of the Poissonian distribution of counts manifests itself strongly.
		The data analysis methods and techniques used to deal with these effects are well known and constitute the standard set of tools in X-ray astronomy.
		A detailed account of these effects and others (e.g.\ confusion with extended sources) is beyond the scope of this paper.

		\subsection{High-redshift AGN} \label{ssec:HighzAGN}

		The density of AGN at high-redshifts is of particular interest because it can place constraints on the formation scenarios of first supermassive black holes and, hence, on cosmological models \citep{Brandt2005}.
		Their numbers in the existing surveys, including those used to produce the XLF models, are very limited.
		Indeed, the highest redshift bin in the AGN sample of H05 was located at $z=3.2-4.8$ and contained  $17$ objects.
		The sample of \citet{Miyaji2000} contained $25$ AGN in a somewhat wider redshift interval of $2.3-4.6$, and the sample of \citet{Ebrero2009} had no AGN with $z>3$.
		Moreover, these samples are not entirely independent because they were obtained from overlapping sets of deep surveys.
		There is only a handful of $z>5$ AGN currently known in X-rays \citep[e.g.][]{Civano2011}.
		Due to low numbers of distant AGN, the XLF at high redshifts is poorly constrained.
		As demonstrated below, eRASS  will significantly enhance the statistics of high-redshift objects. 
		
\begin{figure} 
  	 			\resizebox{\hsize}{!}{\includegraphics{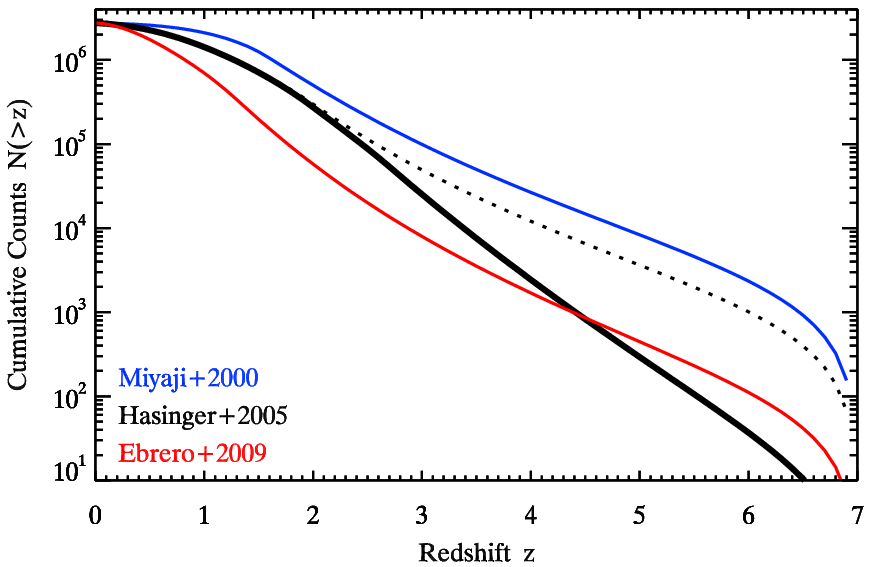}}  
				\caption{\label{fig:N_z_high}%
					Numbers of high-redshift AGN, $N(>\!z)$, expected in the soft band for the extragalactic sky after four years.
					The thick solid black curve shows the prediction based on the default XLF model.
					Also shown are predictions for the XLF model of \citet[][Table~3]{Miyaji2000}, \citet[][Table~5, without the exponential redshift cutoff, dotted curve]{Hasinger2005}, and \citet{Ebrero2009}.
					To obtain these curves we integrated the XLFs to the highest redshift of $z=7$.
					All curves are rescaled to match the average source density computed with our default model (Sect.~\ref{sec:LogNLogS}).
					}
			\end{figure}

		Our poor knowledge of the AGN XLF at high-redshifts limits our ability to accurately predict numbers of high-redshift AGN in the eRASS.  
		To estimate the range of uncertainties we calculated their numbers in the soft band using several different XLF models.
		The resulting cumulative number counts  are shown in Fig.~\ref{fig:N_z_high}.
		For the purpose of this comparison, the curves were rescaled to reproduce the same number density of AGN as the arithmetic mean of our number density map introduced in Sect.~\ref{sec:LogNLogS}.
		The correction factors in the soft band are $1.33$ and $1.32$ for H05 with and without redshift cutoff, respectively, $1.15$ for \citet[][Table~3]{Miyaji2000} and $1.88$ for \citet{Ebrero2009}.

		
		As one can see from the Fig.~\ref{fig:N_z_high}, the number of high-redshift objects in the extragalactic sky sample is uncertain by 
		more than an order of magnitude, different predictions ranging for $z>3$ from $\sim10^4$ to $\sim10^5$.
		For $z>4$ and $z>5$ the numbers vary from $\sim2\,000$ to $\sim30\,000$ and from $\sim300$ to $\sim9\,000$.
		The exponential redshift cutoff of the H05 XLF (cf. solid and dotted black curves) has a significant effect on the numbers of high-redshift sources bringing it close to the prediction based on the XLF of \citet[][red curve]{Ebrero2009}.
		On the other hand, the prediction based on the XLF of H05 without a cutoff (dotted black curve) is close to that of \citet[][blue curve]{Miyaji2000}.
		This large discrepancy of different XLF at high-redshift was previously pointed out by \citet{Brusa2011}.
		
		For the number of objects in the $z=3-5$ redshift range our predictions vary from $\sim8\,000$ to $\sim90\,000$, the default XLF giving $\sim30\,000$ objects; without the exponential redshift cutoff this number is increased by almost a factor of two.
		
		According to our default hard-band XLF, there will be $\sim4$ detected AGN in the hard band for the redshift of $z\gtrsim3.5$.
		However, the discrepancy between different hard-band XLFs is also large, the predictions ranging from $\sim2$ \citep[LADE model of Table~4]{Aird2010} to $\sim200$ \citep{LaFranca2005} AGN for $z\gtrsim3.5$.
		
		The density of high-redshift objects will be higher in the ecliptic poles (Fig.~\ref{fig:N_z} and Table~\ref{tab:Survey}).
		For the default soft-band XLF, there will be $1$ high-redshift ($z\gtrsim5$) AGN every $\sim5\;\mathrm{deg^2}$, which results in $\sim17$ objects in total.
		Without the exponential redshift cutoff this number is an order of magnitude larger. 
		For the other rescaled XLFs from Fig.~\ref{fig:N_z_high} the number of objects varies between $\sim20$ and $\sim600$.
		Obviously, the higher source density and smaller area will facilitate the search for high-redshift objects in the pole regions.
		
		\section{Redshift determination with the iron K$\alpha$ line} \label{sec:Fe}
		
			
				\begin{figure} 
					\resizebox{\hsize}{!}{\includegraphics{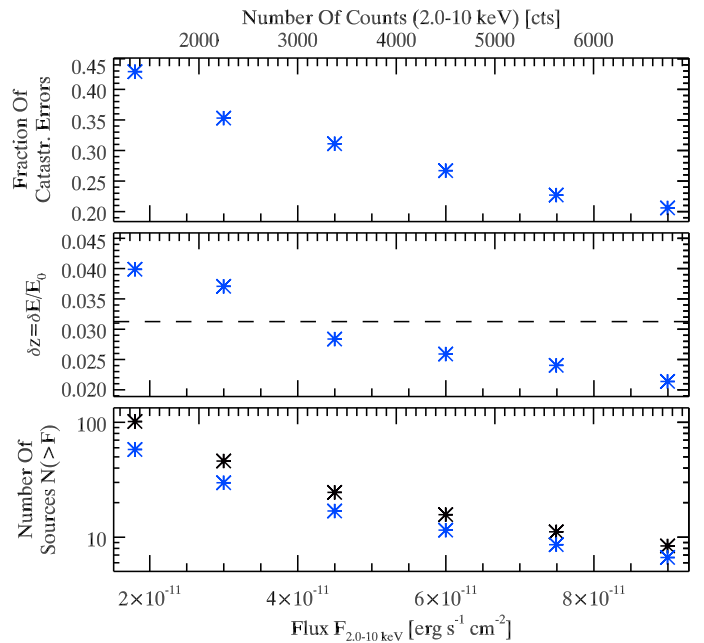}}  
					\caption{\label{fig:Fe}%
						Feasibility of using the iron $K\alpha$ line for the redshift determination at $z=0$.
						See Sect.~\ref{sec:Fe} for details.
						\emph{Top}: Fraction of catastrophic failures.
						\emph{Middle}: Accuracy of the redshift determination. The black dashed line corresponds to the energy resolution of eROSITA.
						\emph{Bottom}: Number of sources for which the redshift can be determined with the accuracy shown in the middle panel, the catastrophic failures excluded (black points show only the $\log N - \log S$).
						}
				\end{figure}

		The strong K$\alpha$ line of iron at $\approx6.4\;\mathrm{keV}$ in the spectra of AGN in principle opens the possibility of determining redshifts by means of X-ray spectroscopy.
		Below, we investigate this possibility for the parameters of eRASS characteristics of the eROSITA telescope. 
			
		We assumed that the continuum spectrum is described by an absorbed power-law with a photon index of $\Gamma=1.9$ and $N_\mathrm{H}=6\times10^{20}$ cm$^{-2}$.
		The shape of the iron K$\alpha$ line may be fairly complex and typically includes narrow and broad components,
		with centroids depending on the ionization state (e.g. $6.4\;\mathrm{keV}$ for the neutral component), and may be complicated by other features
		(e.g. the $7.11\;\mathrm{keV}$ absorption edge; \citealp[e.g.{}][]{Gilli1999,Nandra2007,Corral2008,Shu2010,Krumpe2010a,Chaudhary2012}).
		However, as the final result of this calculation turn out to be somewhat negative, we ignored this complexity and used a simple model%
			\footnote{
				In the flux range of interest, more complex models generally lead to larger fraction of catastrophic failures.
			},
		consisting of a single Gaussian line at $6.4\;\mathrm{keV}$ with an intrinsic width of $\sigma_\mathrm{Fe}=200\;\mathrm{eV}$ and an equivalent width of $150\;\mathrm{eV}$ (rest-frame values).

		To investigate the detectability of the iron K$\alpha$ line in the spectra of eRASS AGN, we performed the following simulations:
		We chose a number of flux values in the $10^{-11}-10^{-10}\;\mathrm{erg\;s^{-1}\,cm^{-2}}$ flux range.
		We fixed the redshift and for each flux value simulated $1\,000$ spectra using the \texttt{phabs(zpowerlw + zgauss)} model in XSPEC \citep[version 12.7.0,][]{XSPEC}.
		Each spectrum was fit with the same model.
		In the fit, the parameters $\sigma_\mathrm{Fe}$ and $N_\mathrm{H}$ were fixed, the initial values of other parameters were set at their simulated values.
		After $1\,000$ spectra were simulated, the distribution of the best values of the redshift was analyzed.
		It was fit with a Gaussian distribution, then the  points outside the $\pm3\sigma$ range were marked as catastrophic failures and clipped out, and the distribution was fit by a Gaussian again. The newly obtained width of the Gaussian  determines the accuracy of the redshift determination $\sigma_z$.
		The catastrophic error fraction was then recomputed as a fraction of objects outside~$\pm3\sigma_z$.
			
		Our results for the redshift $z=0$ are shown in Fig.~\ref{fig:Fe}.
		As one can see from the plot, even at considerably large number of counts, $\sim1\,500$ in the hard band, corresponding to a flux of $\sim2\times10^{-11}\;\mathrm{erg\;s^{-1}\,cm^{-2}}$, the fraction of catastrophic errors  is still large, $\sim40\,\%$.
		This is caused by the steep decrease of the eROSITA efficiency curve with energy, by more than an order of magnitude between 2 and $6\;\mathrm{keV}$.
		As a result, even at large total number of counts, the number of counts at $\sim6\;\mathrm{keV}$, is too small for a reliable line detection in the flux range of interest.
			
		From the  middle panel of the Fig.\ref{fig:Fe}, the accuracy of the redshift measurements for the remaining $\sim60\,\%$ of objects  may seem to be reasonably good, $\delta z\la 0.05$.
		Obviously, this is a result of its definition, which relies on excluding catastrophic failures.
		This definition works well when the there are few catastrophic failures.
		However, the effect of small $\delta z$ is nullified when the fraction of catastrophic failures is large.
		Furthermore, the numbers of objects in this flux range is of the order of $100$ on the entire extragalactic sky, which is too small to be useful.
		The majority, if not all of these bright objects, will be previously known AGN with known redshifts.
			
		The increase of the effective area toward low energies could improve the situation at higher redshifts.
		For a $z\approx2$ object, for example, the observed energy of the iron K$\alpha$ line would fall near the peak of the eROSITA sensitivity and would lower the lowest flux required for a reliable redshift determination using the iron K$\alpha$ line to $\sim10^{-12}\;\mathrm{erg\;s^{-1}\,cm^{-2}}$.
		Unfortunately, the relatively small number of $z\approx2$ objects (a few hundreds) and their low fluxes negate the advantage given by the larger effective area at low energies.
		However, we note that a spectral analysis would still be possible for the sources of known redshift, and that one can still use spectral stacking analysis \citep[e.g.{}][]{Chaudhary2010} to study the average properties of the iron K$\alpha$ line of AGN.

		\section{Transient events and flux variability} \label{sec:Var}
	
		In the four-year survey, eROSITA will scan the whole sky eight times (one full-sky survey per half year).
		The telescope rotates around an axis pointing either directly towards the Sun or with some offset (see Sect.~\ref{sec:Sens})
		and will complete one full circle on the sky every four hours \citep{eROSITA}.
		The plane of the scan rotates with an average rate of one degree per day.
		With this scan geometry, a point on the equator will be scanned $\approx6$ times every half year, separated by $\approx4$ hours.
		The number of consecutive scans per one survey increases with latitude as $\propto \cos^{-1}(\delta)$,  the poles being scanned continuously every four hours during the entire duration of the survey.
		This scan pattern defines two different sampling rates: $\Delta t_1 = 0.5\;\mathrm{years}$ and  $\Delta t_2 = 4\;\mathrm{hours}$, corresponding to frequencies of $\approx6\times10^{-8}$ and $\approx7\times10^{-5}\;\mathrm{Hz}$. 
		
		For one full-sky survey, the average exposure time is $\approx320\;\mathrm{sec}$ %
			\footnote{
				For this calculation it is more appropriate to assume  an observing efficiency of $100\,\%$.\label{fn:Var}
			}.
		At this exposure time, $5$ counts correspond to the flux  of $\approx2\times10^{-14}$ and $\approx4\times10^{-13}\;\mathrm{erg\;s^{-1}\,cm^{-2}}$ in the soft and hard band.
		These numbers define the sensitivity of eRASS to the events (e.g. flares) occurring on the half-year timescale.
		With eight measurements it is also possible to estimate the rms variability on the corresponding timescale.
		To estimate the sensitivity, we took into account that the sample variance is distributed as $\sigma^2\chi^2_{\mathcal{N}-1}/(\mathcal{N}-1)$, where $\sigma$ is the sample rms, $\mathcal{N}$ is the number of points and $\chi^2_{\mathcal{N}-1}$ is the $\chi^2$-distribution with $\mathcal{N}-1$ degrees of freedom.
		Therefore the $1\sigma$ error of the fractional $rms^2$ determination is $\delta(rms^2)\approx\sqrt{2/(\mathcal{N}-1)}\times\left(S/N\right)^{-2}$, where $S/N$ is the signal-to-noise ratio.
		As an estimate of the sensitivity-to-source flux variability, we took the square root of this expression and obtained $\delta(rms)\approx0.73\times \left(S/N\right)^{-1}$.
		Thus, for a $2\times10^{-13}\;\mathrm{erg\;s^{-1}\,cm^{-2}}$ soft-band source, a fractional variability of $rms\sim20\,\%$  will be detected with a $2\sigma$ confidence.
		On the extragalactic sky, about $35\,000$ sources are above this flux threshold.
			 			
	 	At the average exposure time in a single scan ($\approx32\;\mathrm{sec}$), $5$ counts correspond to a flux of $\approx2\times10^{-13}\;\mathrm{erg\;s^{-1}\,cm^{-2}}$ in the soft band.
	 	There are about $39\,000$ sources on the extragalactic sky above this flux level.
	 	This also defines the sensitivity of the eRASS to transient events on an approximate timescale of hours.
	 	Except for the sources in the polar regions, aperiodic variability on these timescales can be measured only for a small number  of sources.

		
			\begin{figure} 
  	 			\resizebox{\hsize}{!}{\includegraphics{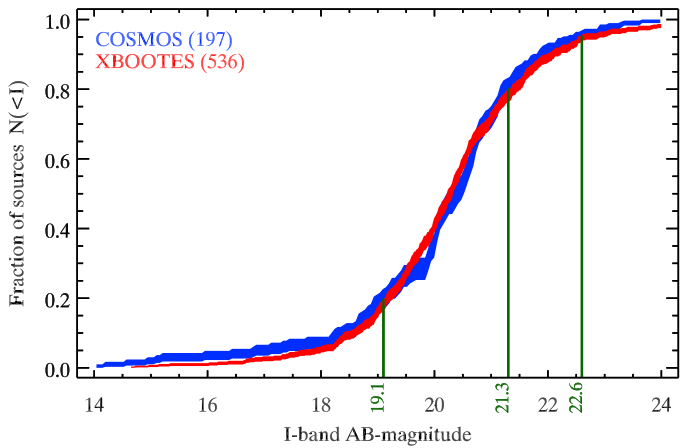}}  
				\caption{\label{fig:Opt}%
					Cumulative $I$-band AB-magnitude distribution of AGN in the COSMOS and XBOOTES fields with the $0.5-2.0\;\mathrm{keV}$ flux exceeding the four-year eRASS detection threshold.
					The thickness of the curves represents the standard deviation of a binomial distribution.
					The vertical lines show the photometric sensitivities of the SDSS ($21.3\;\mathrm{mag}$) and Pan-STARRS PS1 ($22.6\;\mathrm{mag}$) and the magnitude limit for SDSS spectroscopy ($19.1\;\mathrm{mag}$).
					}
			\end{figure}

		\section{Optical counterparts} \label{sec:Opt}
		
		To fully explore the potential of the eRASS, extensive optical coverage will be critical.
		The primary, but not sole goal of this coverage will be to provide identifications and redshift information.
		A detailed discussion of the feasibility and possible strategies of the optical support of the eRASS and its synergies with various ongoing and future optical surveys is beyond the scope of this paper and is currently extensively discussied in the eROSITA collaboration.
		In this Section we investigate the expected optical magnitude distribution of the eRASS AGN. 
		To this end, we used results of the XMM-COSMOS \citep{XMMCOSMOS} and XBootes \citep{Murray2005} surveys.
		For the COSMOS field, we used the results of \citet{Brusa2010}, who cross-correlated the original XMM-COSMOS catalog of X-ray sources of \citet{Cappelluti2009} with the data of the optical survey of the COSMOS field by \citet{Capak2007}.
		From these data we selected sources with a $0.5-2.0\;\mathrm{keV}$ flux, which exceeds the eRASS four-year detection threshold
		(214 out of 1710 sources fulfill the condition  $S_{0.5-2.0\;\mathrm{keV}}\geq10^{-14}\;\mathrm{erg\;s^{-1}\,cm^{-2}}$),
		which had a high identification reliability (204 out of 214), 
		and were not brighter than $I_\mathrm{AB}=14.0\;\mathrm{mag}$ (197 out of 204).
		This selection resulted in a sample of 197 sources for which we obtained the $I$-band AB-magnitude distribution. 
		
		We similarly analyzed the XBootes field, cross-correlating the X-ray and optical catalogs for this field \citep{Kenter2005,Brand2006}.
		We selected point-like sources ($\mathrm{class}\geq0.50$) with $S_{0.5-2.0\;\mathrm{keV}}\geq10^{-14}\;\mathrm{erg\;s^{-1}\,cm^{-2}}$ (565 out of 3213),
		for which an optical counterpart was found ($\mathrm{St}=1$) (565 out of 565) with a high probability of true identification ($\mathrm{Popt}\geq0.95$) (561 out of 565) and an optical flux of $I_\mathrm{AB}>14.0\;\mathrm{mag}$ (536 out of 561).
		We thus selected 536 X-ray sources, for which we computed the cumulative $I$-band magnitude distribution, converting the Vega magnitudes to AB-magnitudes with the conversion factor from \citet[Table~1]{Blanton2007}: m$_\mathrm{AB}$ = m$_\mathrm{Vega}$ + m$_\mathrm{AB}$(Vega) with m$_\mathrm{AB}$(Vega) = 0.45 for the $I$-band.
 
		These $I$-band magnitude distributions for the COSMOS and Bootes fields are plotted in the Fig.\ref{fig:Opt}.
		They show a good agreement between the results for the two different fields, meaning that we have a very good knowledge of the expected magnitude distribution of sources at bright X-ray fluxes.
		Comparing this distribution with the limiting magnitude of the Sloan Digital Sky Survey in the $i$-band, $21.3\;\mathrm{mag}$  at the $95\,\%$ completeness \citep{SDSSDR7}, we conclude that about $\approx80\,\%$ of the eRASS AGN in the SDSS sky will have optical counterparts.
		Taking into account the sky area covered by SDSS, $\approx14\,500\;\mathrm{deg^2}$, we estimate that about $\sim1/3$ of eRASS objects will have an optical counterpart in the SDSS photometric catalog.
		About $\approx20\,\%$ of objects will be brighter than the spectroscopic limit of the SDSS, $i=19.1\;\mathrm{mag}$ for quasars at $z<3$ \citep{Richards2002}, that is, some fraction of these objects will have SDSS spectra.
		
		Repeating this analyzsis for the half-year sensitivity of eRASS, we expecte that almost all eRASS AGN in the SDSS sky will have optical counterparts.
		
		One can see from Fig.~\ref{fig:Opt} that approximately $95\,\%$ of eRASS AGN will be brighter than $I_\mathrm{AB}=22.5\;\mathrm{mag}$ ($R_\mathrm{AB}\approx23.0\;\mathrm{mag}$).
		The Pan-STARRS PS1 $3\pi$ survey will exceed this depth with its expected sensitivity of $\approx22.6\;\mathrm{mag}$ in one visit \citep{Chambers2006}.
		The three year PS1 sensitivity in the $I$-band will reach $\approx23.9\;\mathrm{mag}$ and will cover virtually all eRASS objects in the field of the $3\pi$ survey .


		\section{Summary and conclusions} 
	
		We computed various statistical characteristics of the expected eRASS AGN sample, including their luminosity- and redshift distributions, and the magnitude distributions of their optical counterparts.
	
		The eROSITA all-sky survey will produce an unprecedented sample of about $3$~million X-ray selected AGN.
		With a median redshift of $z\approx1$, approximately $40\,\%$ of the eRASS objects will be located between redshifts $z=1$ and $z=2$ (Fig.~\ref{fig:N_z}).
		We predict that about $10^4-10^5$ AGN beyond redshift $z=3$ and about $2\,000-30\,000$ AGN beyond redshift $z=4$, the exact numbers depending on the behavior of the AGN XLF in the high-redshift and high-luminosity regimes (Fig.~\ref{fig:N_Lz}).
		
		The eRASS AGN sample will open the possibility of studying the growth of supermassive black holes, their co-evolution with host galaxies and dark matter halos, and their relation to the large-scale structure to unprecedented detail, and potentially, it may also help to constrain cosmological parameters \citep{Kolodzig2013,Huetsi2013}.
		Importantly, it will permit one to conduct these studies beyond redshift $z=1$, which is poorly covered by the current optical surveys.
		
		To fully exploit this potential of eRASS, an extensive optical support will be critical.
		One of the main goals of optical follow-up will be to provide redshifts for eRASS AGN, but its importance will reach far beyond this, including, for example,  studies of the co-evolution of supermassive black holes and their host galaxies
		(see \citealp{Kolodzig2013} for a discussion of the different goals of optical follow-ups and their requirements and prospects).
		With the capabilities of the currently available facilities and their time allocation strategies, measuring of optical  spectra for the entire sample of the $\sim3$~million objects does not appear to be achievable on realistic timescales.
		However upcoming hardware and survey programs and proposals, for instance, 4MOST \citep{deJong2012} and WEAVE \citep{WEAVE}, can make this task more realistic, especially for some limited areas of sky.
		Furthermore, introducing mutliband photometry and other improvements of the photometric redshift measurement techniques will make determining of photometric redshifts  for large samples of eRASS AGN possible \citep{Salvato2011,Saglia2012}.

		
 		\begin{acknowledgements}
  			A. Kolodzig acknowledges support from and participation in the International Max-Planck Research School (IMPRS) on Astrophysics at the Ludwig-Maximilians University of Munich (LMU).
  			M. Brusa acknowledges support from the FP7 Career Integration Grant 'eEASy: eROSITA Euclid AGN Synergies' (CIG 321913).
  			The research made use of grant NSh-5603.2012.2 and programs P-19 and OFN-17 of the Russian Academy of Sciences.
  			
  			We thank Jan Robrade for the exposure-time map of eROSITA and additional calculations, Peter Friedrich for the details about the PSF of eROSITA, Thomas Boller for the details about the eROSITA background calculation, Takamitsu Miyaji for discussions of AGN XLF, Andrea Merloni for useful discussions, and Peter Predehl for background information about eROSITA.
 		\end{acknowledgements} 
		
			\bibliographystyle{aa}

\end{document}